\documentclass[11pt,letterpaper]{article}
\usepackage[top=0.8in,left=0.8in,footskip=-0.25in]{geometry}

\usepackage{amsmath,amssymb}

\usepackage{annotate-equations}

\usepackage{changepage}

\usepackage{textcomp,marvosym}

\usepackage{cite}

\usepackage{nameref,hyperref}
\makeatletter
\providecommand{\leftsquigarrow}{%
  \mathrel{\mathpalette\reflect@squig\relax}%
}
\newcommand{\reflect@squig}[2]{%
  \reflectbox{$\m@th#1\rightsquigarrow$}%
}
\makeatother
\usepackage{dsfont}
\usepackage{mathtools} 
\usepackage[final]{changes}
\usepackage[right]{lineno}

\usepackage[nopatch=eqnum]{microtype}
\DisableLigatures[f]{encoding = *, family = * }

\usepackage{array}

\newcolumntype{+}{!{\vrule width 2pt}}

\newlength\savedwidth

\usepackage{setspace} 

\usepackage[aboveskip=1pt,labelfont=bf,labelsep=period,
  singlelinecheck=off]{caption}

\makeatletter
\renewcommand{\@biblabel}[1]{\quad#1.}
\makeatother

\usepackage{lastpage,fancyhdr,graphicx}
\usepackage{epstopdf}
\fancyheadoffset[L]{2.25in}
\fancyfootoffset[L]{2.25in}
\def\dd{\textrm{d}}

\graphicspath{{FIG/}}
\begin{document}

\begin{flushleft}
{\Large
\textbf{Reliable ligand discrimination in stochastic multistep kinetic proofreading:
First passage time vs. product counting strategies} 
}
\newline
\\
Xiangting Li\textsuperscript{1},
Tom Chou\textsuperscript{1,2*},
\\
\bigskip
\textbf{1} Department of Computational Medicine, University of California, Los Angeles, CA 90095, USA.
\\
\textbf{2} Department of Mathematics, University of California, Los Angeles, CA 90095, USA.
\\
\bigskip

* tomchou@ucla.edu

\end{flushleft}
\section*{Abstract}
Cellular signaling, crucial for biological processes like immune
response and homeostasis, relies on specificity and fidelity in signal
transduction to accurately respond to stimuli amidst biological noise.
Kinetic proofreading (KPR) is a key mechanism enhancing signaling
specificity through time-delayed steps, although its effectiveness is
debated due to intrinsic noise potentially reducing signal fidelity.
In this study, we reformulate the theory of kinetic proofreading (KPR)
by convolving multiple intermediate states into a single state and
then define an overall ``processing'' time required to traverse these
states. This simplification allows us to succinctly describe kinetic
proofreading in terms of a single waiting time parameter, facilitating
a more direct evaluation and comparison of KPR performance across
different biological contexts such as DNA replication and T cell
receptor (TCR) signaling. We find that loss of fidelity for longer
proofreading steps relies on the specific strategy of information
extraction and show that in the first-passage time (FPT)
discrimination strategy, longer proofreading steps can exponentially
improve the accuracy of KPR at the cost of speed. Thus, KPR can still
be an effective discrimination mechanism in the high noise
regime. However, in a product concentration-based discrimination
strategy, longer proofreading steps do not necessarily lead to an
increase in performance. However, by introducing activation thresholds
on product concentrations, can we decompose the product-based strategy
into a series of FPT-based strategies to \added{better resolve the
  subtleties of KPR-mediated product discrimination.}
%
%
%
Our findings underscore the importance of understanding KPR in the
context of how information is extracted and processed in the cell.


\section*{Author summary}
Kinetic proofreading (KPR) is mechanism often employed by cells to
enhance specificity of ligand-receptor.  However, the performance of
kinetic proofreading may be hampered by noise and a low
signal-to-noise ratio. By consolidating multiple kinetic proofreading
steps into a single state and assigning an associated waiting, or
``processing time,'' we developed an analytic approach to quantify the
performance of KPR in different biological contexts. Despite a
trade-off between speed and accuracy inherent to a first-passage time
KPR strategy, we show that a signaling molecule-based discrimination
strategy can enhance the performance benefits of KPR.  We further
decompose the product-based discrimination strategy into a set of
first-passage times to different thresholds of signaling molecules
produced. Through this decomposition, we find that a threshold that
adjusts dynamically throughout the recognition process depends on the
duration of the process. We propose that this more nuanced
product-based KPR-mediated recognition process can be realized
biologically. The precise structural basis for a dynamic threshold
merits further experimental exploration, as it may hold significant
implications for understanding biological mechanisms of information
transmission at a molecular level.


\section*{\label{sec:intro}Introduction} Various cellular
processes require a high degree of specificity in order to function
properly, including DNA replication, gene expression, and cellular
signaling. The degree of specificity observed is often hard to justify
by a simple binding-affinity argument, the specificity of which is
proportional to $\exp(- \Delta\Delta G/ RT)$, where $\Delta \Delta G$
is the difference in free energy between the correct and incorrect
ligands \cite{Hopfield1974oct}. For example, the estimated error
probability per nucleotide in DNA replication is estimated to be
$10^{-9}$ \cite{Kimura1974July}, but the net free energy difference
between mismatched and matched base pairs is only $0.5~
\text{kcal}/\text{mol}$ \cite{Petruska1988September}, suggesting the
theoretical error rate would be $\sim 10^{-3}$.  Similarly, T cells
need to specifically distinguish self-antigens from mutant
self-antigens, also known as neoantigens, which can differ by only one
or a few amino-acids \cite{Davis2021December}.

\begin{figure}[htb]
  \centering
  \includegraphics[width=6.5in]{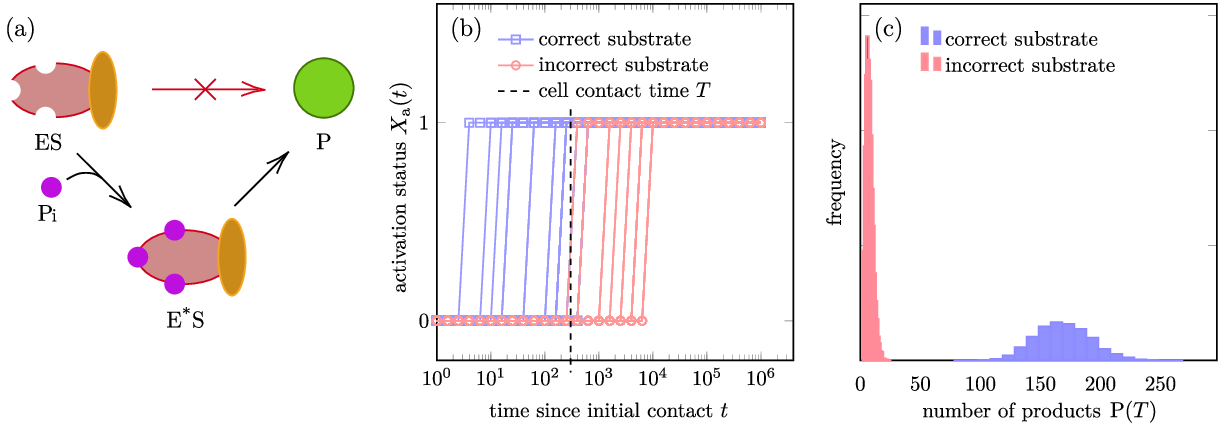}
  \vspace{4mm}
    \caption{Schematic of the KPR process and different strategies of
      interpreting the output. (a) The complex of enzyme and substrate
      ${\rm ES}$ alone cannot produce the final product ${\rm P}$. It
      has to undergo a number of proofreading steps, represented here
      by phosphorylation (${\rm Pi}$), before the activated state
      ${\rm E^* S}$ can produce the final product. (b) The FPT-based
      discrimination strategy, simply reaching the activated state
      ${\rm E^* S}$ is interpreted as the output.  $X_{\rm a}(t)=0$ if
      the system has not reached the activated state by time $t$, and
      $X_{\rm a}(t)=1$ otherwise. At $t=0$, the reaction starts with
      the enzyme in the free state ${\rm E}$. The dashed vertical line
      represents the termination of entire process, \textit{e.g.}, the
      time $T$ at which the T cell and antigen presenting cell (APC)
      separate.  $X_a(t)=1$ for some $t<T$ is interpreted as a
      positive response.  (c) The product-based discrimination
      strategy. In this strategy, the number of product molecules
      ${\rm P}(T)$ produced within a given time $T$ is interpreted as
      the output. Due to the intrinsic noise, the number of product
      molecules is a random variable and their distribution is shown
      for correct and incorrect ligands.}
    \label{fig:schematic}
  \end{figure}

Kinetic proofreading (KPR)
\cite{Hopfield1974oct,Ninio1975jul,McKeithan1995may} typically denotes
a chemical reaction mechanism that can significantly increase the
specificity towards a desired ligand against competing ligands. In the
KPR context, ``proofreading'' is accomplished by introducing
additional irreversible, energy-consuming kinetic steps which
individually may not distinguish desired ligands from undesired ones.
However, these steps impart a delay to final product release allowing
for ``resetting'' of the process and an overall lower final error rate
\added{(a multistep KPR mechanism is illustrated in
  Fig.~\ref{fig:schematic}(a) below and mathematical details are
  discussed in the Materials and Methods).}

First proposed by Hopfield to explain the high specificity of
DNA/protein synthesis \cite{Hopfield1974oct}, the KPR mechanisms have
been invoked to explain other biological processes such as T cell
receptor (TCR) signaling \cite{McKeithan1995may} and microtubule
growth \cite{Murugan2012jul}.  These first treatments of KPR described
it within the steady-state limit of deterministic mass-action models,
comparing the steady-state fluxes of the correct and incorrect
  product formation. Reactions inside the cell, however, are often
between small numbers of molecules and are thus stochastic. Stochastic
aspects of KPR have also been considered, emphasizing the statistics
of first passage times (FPT) to product formation
\cite{Lipniacki2008sep,Bel2009dec,Morgan2023aug}.

Recently, Kirby and Zilman reported that adding more kinetic
proofreading steps almost always decreases the signal-to-noise ratio
(SNR) defined by the ratio of the mean to the standard deviation of
the output signal \added{(the number of signaling molecules
  produced)}, suggesting that KPR is not an optimal strategy for
information processing due to noise \cite{Kirby2023may}.  However, TCR
signaling and T cell activation, the context that Kirby and Zilman
describe, is a highly specific process that does involve multiple
kinetic proofreading steps, \added{but with adaptive variants} of KPR used to
model TCR signaling
\cite{AltanBonnet2005oct,Lalanne2013may,Tischer2019apr,
  Pettmann2021may,Voisinne2022aug}.

In this paper, we reconcile the apparent contradiction between the
high specificity of TCR signaling and the low SNR of a longer-chain
KPR process. The key theoretical insight involves convolving the
multiple intermediate irreversible steps into a single equivalent
state in which the system stays for time $\tau$. Instead of explicitly
treating a series of sequential states, we define a single, equivalent
waiting time or ``processing'' time $\tau$. This simple reduction
reveals intriguing insights, allows us to analytically and
systematically explore different biological contexts of KPR, and
provides an easier framework on which to test different metrics of KPR
performance.

We show that the apparent contradiction arises from different
strategies of determining whether a final output is correct. In the
FPT-based scenario for both DNA replication and TCR signaling,
arriving at an activated state or a ``product'' state within a given
time is interpreted as the output, as illustrated in
Fig.~\ref{fig:schematic}(b). A longer processing time $\tau$ can
exponentially improve the accuracy of KPR at the cost of speed. The
trade-off between speed and accuracy has been reported in experimental
studies \cite{Johansson2011dec,Savir2013apr,Banerjee2017May}. In an
alternative strategy for TCR signaling implicitly used by Kirby and
Zilman, the maximum SNR, or mutual information between the input and
output arises with just two proofreading steps. However, more
processing steps leads to a decrease in KPR performance.

\added{The alternative discrimination strategy implicit in Kirby's
  metric} is to detect the number of products (\textit{e.g.},
signaling molecules that lead to downstream processes) generated
within a finite time without explicitly resolving the final response
of the cell, \added{as is illustrated in
  Fig.~\ref{fig:schematic}(c)}. Mutual information has been used in
recent studies to quantify the information flow in cellular
decision-making processes
\cite{Cheong2011October,Selimkhanov2014dec,Tang2021February}.  Here,
we introduce mutual information and channel capacity in order to
compare the performance of the two strategies (FPT to a target state
and product counting) on equal footing. We also construct a
decomposition of the product-based strategy into a series of FPT-based
strategies with different product molecule thresholds and conclude
that the product detection strategy is equivalent to a strategy that
dynamically adjusts the threshold according to the duration of the
process.  This dynamic thresholding strategy can be shown to be more
robust to fluctuations over the duration of the reaction.  The
effectiveness of this strategy can be attributed to an additional
layer of proofreading.

Our analysis and findings present a unified framework for analyzing
KPR under different biological scenarios. We also highlight the
importance of understanding how different strategies of information
extraction can affect the performance and parameter tuning of KPR.

\section*{Materials and methods}
In this section, we first describe the general model of KPR and then
apply it to two specific biological contexts, DNA replication and TCR
signaling.

\subsection*{\label{sec:model}Model settings}
In the conventional setting of KPR, the complex ${\rm E}^{(0)}{\rm S}$
composed of receptor ${\rm E}$ and a ``correct'' substrate (or ligand)
${\rm S}$ forms and dissociates with binding and unbinding rates
$k_{\pm 1}$.  A complex with the ``incorrect'' ligand forms and
dissociates with rates $q_{\pm 1}$. Both complexes can undergo
multiple nonequilibrium transitions or proofreading steps
(\textit{e.g.}, sequential phosphorylation) traversing internal states
$({\rm E}^{(1)}{\rm S}, ..., {\rm E}^{(m-1)}{\rm S})$ before the final
product ${\rm P}$ can be released or produced by the fully activated
state ${\rm E}^*{\rm S}$.

\begin{figure}[htb]
\centering 
\includegraphics[width=4.5in]{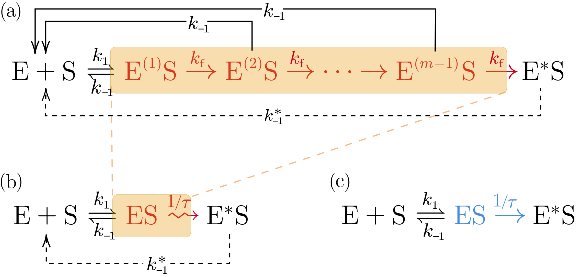}
\vspace{4mm}
\caption{Reaction schemes of different descriptions of the simple
  kinetic proofreading process. (a) The conventional description of
  KPR explicitly incorporating multiple proofreading steps. (b) A
  reduced representation of KPR in which multiple driven steps are
  lumped in a single proofreading step. For comparison, we show the
  classical Michaelis-Menten reaction scheme in (c). In (a) and (b),
  unbinding of ${\rm E}^*{\rm S}$ is not essential, \textit{i.e.}, it
  is possible that $k_{-1}^*=0$, thus marked in dashed lines.
  Additionally, in (a-b), there can be a production step after ${\rm
    E}^*{\rm S}$ is reached, which is not explicitly shown.}
    \label{eq:reaction}
\end{figure}
Each internal state of the complex can dissociate with rate $k_{\rm
  off}$ or proceed to the next step with rate $k_{\rm f}$, as shown in
Fig.~\ref{eq:reaction}(a).  To simplify our subsequent analysis, we
set $k_{\rm off} = k_{-1}$ as in \cite{Kirby2023may}. Additionally, in
the multistep limit ($m \rightarrow \infty$), the processing time
$\tau$ of reaching the final activated state ${\rm E^*S}$ from ${\rm
  ES}$ is fixed to a deterministic value $\tau = m/k_{\rm f}$,
\added{(in which $k_{\rm f}$ scales with $m$ as $m \rightarrow
  \infty$)} as was analyzed by \cite{Bel2009dec}. We can thus simplify
the reaction diagram in Fig.~\ref{eq:reaction}(a) by lumping the
internal states $({\rm E}^{(0)}{\rm S},...,{\rm E}^{(m-1)}{\rm S})$
into a single state ${\rm E}{\rm S}$ as shown in
Fig.~\ref{eq:reaction}(b). Note that because of independence between
activation and dissociation in these steps, the transition of the
aggregated state ${\rm E}{\rm S}$ to the dissociated state ${\rm
  E}+{\rm S}$ is still Markovian with the same rate $k_{-1}$. The
master equation of the simplified model and its relation to the
original multi-step model are discussed in Appendix
\ref{appendix:master-equation}.

As shown in Fig.~\ref{eq:reaction}(c), our simplified model has a
similar structure to the classical Michaelis-Menten reaction scheme.
However, our simplified model captures a crucial element of the KPR
mechanism. \added{In the Michaelis-Menten scheme, the waiting time
  $\tau$ in the complex state ${\rm E}{\rm S}$ before converting to
  product is exponentially distributed.  In the KPR scheme, the
  waiting time $\tau$ in the state ${\rm E}{\rm S}$ before converting
  to the activated state ${\rm E}^*{\rm S}$ is assumed to be
  non-exponentially distributed. An exponentially distributed waiting
  time reflects a memoryless process in which the evolution of the
  system depends only on the current state.  This memoryless property
  is the defining feature of a Markov process. By contrast, memory in
  the KPR reaction process results in a non-exponential distribution
  of the processing time $\tau$. In the simplification of the KPR we
  are considering, the processing time is fixed to the value $\tau$,
  \textit{i.e.}, its distribution is a Dirac delta function at $\tau$.
  For example, the waiting time of activation does not depend only on
  the current state, but also on the time elapsed since the initial
  formation of the complex, which happened in the past.}  This
non-Markovian step acts as a memory of the system or a clock that
keeps track of the time elapsed since the initial formation of the
complex, \added{and can be achieved by, \textit{e.g.}, tracking the
  phosphorylation state of the complex ${\rm ES}$}.  The biological
context in which KPR operates will be a determining factor in KPR
model structure and in the specification the most appropriate
performance metric.

\subsection*{DNA replication setting}
First, we consider the DNA replication scenario, specifically a single
nucleotide incorporation step catalyzed by DNA polymerase ($E$). We
track the system starting from an initial state where the enzyme is
free. The correct substrate ${\rm S}$ refers to the complementary
nucleotide to the template strand, while the incorrect substrate ${\rm
S}'$ refers to the other three nucleotides. The enzyme can bind to
either substrate with rates $k_1$ and $q_1$, respectively. The enzyme
can also unbind from either substrate with rates $k_{-1}$ and
$q_{-1}$.

The reaction diagram in Fig.~\ref{eq:reaction}(b) is translated as
\begin{equation}
\begin{aligned}
    {\rm E} + {\rm S} &\underset{k_{-1}}{\overset{k_1}{\rightleftharpoons}} {\rm E}{\rm S} 
    \stackrel{1/\tau}{\rightsquigarrow}  {\rm E}+ {\rm P},\\
    {\rm E} + {\rm S}'&\underset{q_{-1}}{\overset{q_1}{\rightleftharpoons}} {\rm E}{\rm S}' 
    \stackrel{1/\tau}{\rightsquigarrow} {\rm E} + {\rm P}',
\end{aligned}
    \label{eq:dna-replication}
\end{equation}
where ${\rm S}$ and ${\rm S}'$ denote the correct and incorrect
substrates and ${\rm P}$ and ${\rm P}'$ denote the correct and
incorrect products. For simplicity, we identify the activated state
${\rm E}^*{\rm S}$ of ${\rm E}^*{\rm S'}$ with final products ${\rm E
  + P}$ or ${\rm E + P'}$, referring to incorporation of the correct
or incorrect nucleotide, respectively.  

\begin{figure}
  \centering
    \includegraphics*[width=2.5in]{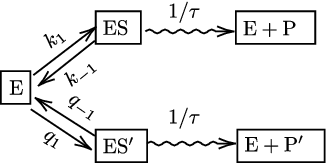}
\vspace{4mm}
    \caption{\label{fig:dna-replication} The simplified model of KPR
      in DNA replication with only one enzyme. Here, the enzyme
      (\textit{e.g.}, DNA polymerase) has three states, namely, free
      $({\rm E})$, bound to correct substrate $({\rm E}{\rm S})$, and
      bound to incorrect substrate $({\rm E}{\rm S}')$. These states
      interconvert with rates specified in the model. When the enzyme
      is bound to substrate, it produces the product (${\rm P}$ or
      ${\rm P}'$) after a waiting time $\tau$.}
\end{figure}

We track the system until either one of the products is produced,
allowing repeated binding and unbinding of the enzyme to the
substrates.  While there can be multiple replication forks in a cell,
we focus on a single DNA polymerase in this study. Thus, we can
represent the stochastic system by a simplified stochastic process
with described by three transient states indicating the status of the
DNA polymerase; namely, unbound polymerase $({\rm E})$, polymerase
bound to correct substrate $({\rm E}{\rm S})$, and polymerase bound to
incorrect substrate $({\rm E}{\rm S}')$. There are also two absorbing
states, namely, correct product $({\rm E}+{\rm P})$ and incorrect
product $({\rm E}+{\rm P}')$. This simplified stochastic scheme is
shown in Fig.~\ref{fig:dna-replication}.

The performance of the KPR process is quantified by the probability
$\mathbb{P}(t_{\rm p} < t_{\rm p'})$ that the FPT $t_{\rm p}$ to the
correct product state $({\rm E}+{\rm P})$ is less than the FPT $t_{\rm
  p'}$ to the wrong product $({\rm E}+{\rm P}')$. By the standard
approach of conditioning on the next state, we find
\begin{equation}
  \begin{aligned}
    \mathbb{P}(t_{\rm p} < t_{\rm p'}) & = \frac{k_1}{k_{1} + q_{1}e^{(k_{-1}- q_{-1})\tau}}, \\
    \mathbb{P}(t_{\rm p} \geq t_{\rm p'}) & = \frac{q_1 e^{(k_{-1}- q_{-1})\tau}}{k_{1} + q_{1}e^{(k_{-1}- q_{-1})\tau}}.
    \end{aligned}
    \label{eq:eta-dna}
\end{equation}
Using the same conditioning approach, we can also calculate the mean
first passage time (MFPT) $\mathbb{E}[t]$ to either product starting
from the free state ${\rm E}$:
\begin{equation}
  \begin{aligned}
  \mathbb{E}[t] &= \frac{\frac{k_1}{k_{-1}}\left(1 - e^{-k_{-1} \tau}\right) + \frac{q_1}{q_{-1}}\left(1 - e^{-q_{-1} \tau}\right) + 1}{k_1 e^{-k_{-1} \tau} + q_1 e^{-q_{-1} \tau}}\\
  &\approx \frac{e^{k_{-1}\tau}}{k_1} \left( \frac{k_1}{k_{-1}}
  + \frac{q_1}{q_{-1}} + 1 \right),
  \end{aligned}
  \label{eq:mfpt-dna}
\end{equation}
where the above approximation holds if $k_1 e^{- k_{-1} \tau} \gg q_1
e^{-q_{-1} \tau}$, which is typically the case in DNA replication, as
$\mathbb{P}(t_{\rm p} < t_{\rm p'})\lesssim 1$.

\subsection*{TCR signaling setting}
Next, we consider proofreading in the TCR signaling context.  \added{T
  cells form contact with an antigen presenting cell (APC) and scans
  the surface of the APC for the presence of a foreign antigen (or
  epitope) $({\rm S})$. The TCR on the T cell membrane can bind to the
  epitope on the APC membrane which initiates a signaling cascade that
  leads to T cell activation. The activated T cell can then produce
  signaling molecules that trigger the immune response. If the APC
  does not carry the foreign antigen corresponding to the TCR, the T
  cell should not be activated within the contact time $T$ and
  disengages from the APC.} There are three features of TCR signaling
that are distinct from the DNA signaling process.  First, the APC is
likely not to carry a foreign antigen, \textit{i.e.}, the correct
substrate $({\rm S})$. Consequently, the correct and incorrect
peptide-MHC complex (pMHC), or ${\rm S}$ and ${\rm S}'$, are not
present at the same time during an encounter, \textit{i.e.}, they do
not compete with each other for TCR binding. Second, the APC and the T
cell have a finite contact duration $T$. The recognition process can
only occur within this time window. Lastly, the activated states
produce identical products, \textit{i.e.}, downstream signals,
regardless of whether the substrate (epitope) is the correct one or
not. Therefore, the cell needs a strategy to discriminate the correct
and incorrect substrates based on the number of products generated
within the contact time $T$.

First, assume that $T$ is fixed as in previous literature
\cite{Morgan2023aug} (we will relax this assumption later on).
\textit{Within the time window $T$}, the reaction diagram is
represented by Eq.~\eqref{eq:multistep}.

\vspace{3mm}

\begin{equation}
  \begin{aligned}
    \tikzmarknode{ini}{{\rm E} + {\rm S}} &
    \underset{k_{-1}}{\overset{k_1}{\rightleftharpoons}} {\rm E}{\rm S} 
    \stackrel{1/\tau}{\rightsquigarrow}
    \tikzmarknode{com}{{\rm E}^*{\rm S}}
    \stackrel{k_{\rm p}}{\rightarrow}P+ {\rm E}^*{\rm S} ,\\
  \end{aligned}
  \label{eq:multistep}
\end{equation}
\tikzset{annotate equations/arrow/.style={->}}
\annotatetwo[yshift=0.7em]{above}{com}{ini}{$k_{-1}^*$} Typically, TCR
recognition is fairly sensitive since a few correct substrates (or
epitopes) ${\rm S}$ on the APC membrane are able to activate the T
cell.  In the following, we will assume that each APC contains only
one type of epitope, ${\rm S}$ or ${\rm S'}$, while there are multiple
TCRs that can bind to the substrate. Mathematically, the roles of
substrate and TCR are equivalent. Tracking the state of the single
substrate gives a similar simple stochastic process as in the DNA
replication setting. This assumption allows us to reduce the number of
possible states in the corresponding stochastic process.

We first consider two possible discrimination strategies, the
FPT-based strategy and the product-based strategy, to quantify 
the output of the TCR recognition process.

\paragraph*{FPT-based scenario.} In this strategy, reaching the
activated state ${\rm E}^* {\rm S}$ within time $T$ is interpreted as
T cell activation. The output of this strategy can be represented
using the FPT to ${\rm E}^* {\rm S}$ state ($t_{\rm a}$) as $X_{\rm a}
= \mathds{1}_{t_{\rm a} < T}$. The state $X_{\rm a} = 1$ denotes an
activated T cell while $X_{\rm a} = 0$ indicates no response by the T
cell. \added{In general, we can define $X_{\rm a}(t) =
\mathds{1}_{t_{\rm a} < t}$, and visualize the output as a binary
signal changing over time, as shown in Fig.~\ref{fig:schematic}(b).}
\added{To justify the FPT strategy biologically, we note that} upon reaching
the activated state, cofactors such as CD4 and CD8 stabilize the
TCR-pMHC interaction, significantly reducing the unbinding rate
$k_{-1}^*$. Therefore, the activated complex can then steadily produce
downstream signals (products) and trigger T cell response. 

\paragraph*{Product-based scenario.} This strategy, which is
implicitly analyzed in \cite{Kirby2023may}, estimates the pMHC-TCR
``affinity'' by counting the number of product or signaling molecules
produced within a given time.  The output of this strategy is the
number of products ${\rm P}(T)$ at time $T$, which takes on values in
$\mathbb{N}$, as is shown in Fig.~\ref{fig:schematic}(c).
Consequently, in this more graded strategy, whether a T cell is
activated is not described by a single, specific criterion.

\added{Different strategies of interpreting the output requires us to
  define performance metrics that can compare different strategies on
  a common mathematical footing. We now define the performance metric
  that can be used to compare our two discrimination scenarios.}

\paragraph*{Performance metrics}
In the case of FPT-based discrimination, it is natural to formulate
the recognition problem as a hypothesis testing problem. We denote the
binary input $\xi=1$ if ${\rm S}$ is present and $\xi = 0$ if ${\rm
  S}'$ is present. The competing hypotheses are then $H_0: \xi =0$ and
$H_a: \xi =1$.  The cell accepts $H_0$ if $X_{\rm a}=0$. There is a
canonical definition of sensitivity and specificity, \textit{i.e.},
the true positive probability $({\rm TPP})$ and true negative
probability $({\rm TNP})$. \added{Given a fixed duration $T$,} varying
the processing time $\tau$ gives a family of binary classifiers
(solutions to the hypothesis testing problem) corresponding to the KPR
process. The receiver-operating characteristic curve (ROC) and the
area under the curve (AUC) can be used to evaluate the overall
performance of this family of classifiers. For a single classifier, we
define the accuracy $\mathcal{A}$ as an average of specificity and
sensitivity:
\begin{equation} \label{eq_acc:accuracy}
  \mathcal{A}\equiv\textrm{accuracy} =
  \frac{1}{2}\textrm{sensitivity} + \frac{1}{2}\textrm{specificity}.
\end{equation}

However, such metric only applies to strategies with binary output,
but not the product-based strategy, as the output ${\rm P}(T)$ is
non-binary. Kirby et al. propose a Fisher linear discriminant metric
($\eta_{\rm FLD}$) based on the consideration of signal-to-noise
ratios \cite{Kirby2023may}. $\eta_{\rm FLD}$ takes on values in $(0,
\infty)$ and does not directly quantify the fidelity of transmission
from the input $\xi$ to the output ${\rm P}(T)$.

In order to compare both strategies on a common footing and represent
the fidelity of discrimination directly, we introduce the mutual
information $\mathcal{I}$ between the input and output, and the
associated channel capacity $C$. The mutual information between two
random variables $\xi$ and $X$ can be defined as \cite{Cover1999}
\begin{equation}
    \mathcal{I}(\xi; X) = S(\xi \otimes X) - S(\xi, X),
    \label{eq:mutual-information}
\end{equation}
where $\xi \otimes X$ is the joint random variable of $\xi$ and $X$,
assuming that $\xi$ and $X$ are independent, while $S(\xi, X)$ is the
joint Shannon entropy of $\xi$ and $X$. The mutual information
$\mathcal{I}(\xi; X)$ relies on the input distribution of
$\xi$. Hence, one can define the channel capacity $C$ as the supremum
of the mutual information over all possible input distributions of
$\xi$ which only depends on the conditional probability distribution
of $X$ given $\xi$,
\begin{equation}
    C(X \mid \xi) = \sup_{\xi} \mathcal{I}(\xi; X).
    \label{eq:channel-capacity}
\end{equation}
There are two advantages of using mutual information and channel
capacity over the Fisher linear discriminant metric $\eta_{\rm
  FLD}$. First, the mutual information is defined for both binary
variables, as in the case of the first-passage time problem, and
continuous variables, as in the case of the product-based
discrimination problem. Second, in the specific scenario of binary
input variables (correct and incorrect substrates), the mutual
information always takes values between 0 and 1 when measured in bits
($\log 2$). A mutual information of 0 means that the distribution of
input and output do not overlap while a value of 1 indicates that the
distributions are identical. \added{Consequently, the channel capacity
  provides a natural way to compare the product-based discrimination
  problem with the FPT problem and to quantify how well the system can
  distinguish correct substrates from incorrect ones.}

In the limit in which the accuracy $\mathcal{A} \rightarrow 1$ with binary
input $\xi$ and output $X$, we note that the mutual information
$\mathcal{I}$ is approximately $\mathcal{A}$ (measured in bits) when
$\mathbb{P}(\xi=0)=\mathbb{P}(\xi=1)=\tfrac{1}{2}$:
\begin{equation}
  \begin{aligned}\label{eq:information_accuracy}
    \mathcal{I}(X_{\rm a}; \xi) \approx 
    \mathcal{A} \textrm{ bits}.
  \end{aligned}
\end{equation}
In this high accuracy limit, the channel capacity $C$ is also close to
the accuracy.

\paragraph*{Stochastic simulations}
In cases where we need to rely on stochastic simulation of KPR
processes, we implement the Gillespie algorithm
\cite{Gillespie1977dec} in \texttt{julia} \cite{julia}.  The Gillespie
algorithm tracks the state transitions of a Markov chain with
exponential waiting times. In order to simulate the KPR process with a
deterministic waiting time $\tau$, we explicitly track all waiting
times at each step and the time elapsed, updating the state by the
smallest waiting time. After the state update, we re-evaluate all the
waiting times and the time elapsed. The implementation is available at
\url{github.com/hsianktin/KPR}.

In order to obtain the mutual information and channel capacity, we
record the FPTs during the simulation, as well as the number of
products produced by each simulation trajectory after a given time
$T$. We simulated $10^4$ trajectories and use the empirical
  distributions of the FPTs and the number of products as surrogates
  for the true distributions. This allows us to compute the
conditional probability distribution of output $X$ given input $\xi$.
Then, mutual information is computed using
Eq.~\eqref{eq:mutual-information}. The channel capacity is obtained by
a numerical optimization procedure with respect to the probability of
$\xi=1$, $\mathbb{P}(\xi=1)$ in the interval $(0,1)$.

The exact parameters used in each simulation will be specified in the
corresponding figures. In general, we set $k_{-1}=1$, $q_{-1}=2$,
indicating that the unbinding rate of an incorrect substrate is only
twice as fast as that of a correct substrate.

\section*{Results}
\subsection*{Long processing time improves maximal accuracy in FPT-based strategy}

\begin{figure}
  \centering
  \includegraphics[width=5.4in]{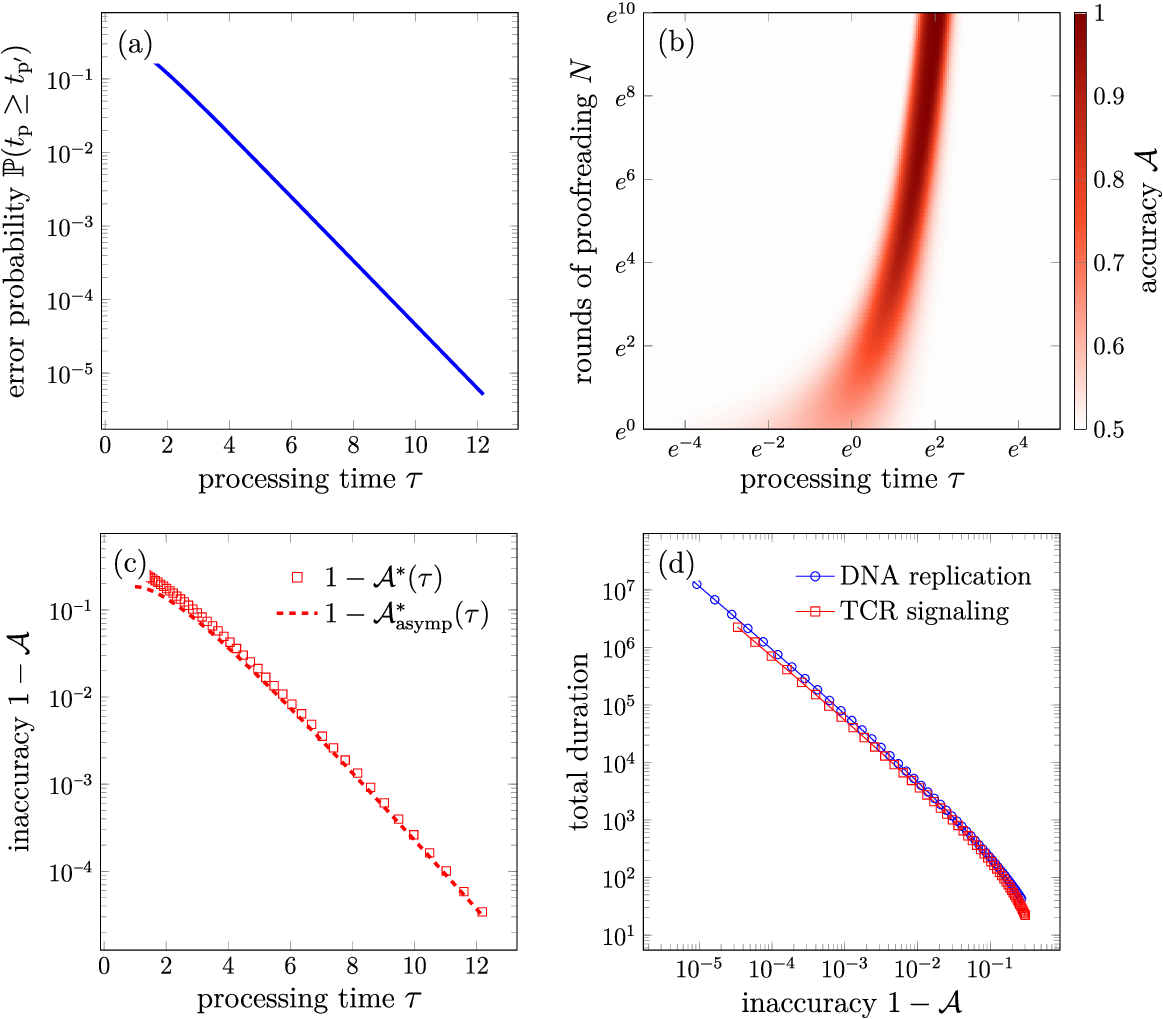}
  \vspace{4mm}
  \caption{Statistics of the FPT-based strategy in the DNA replication
    and TCR recognition scenarios. (a) Replication error probability
    $\mathbb{P}(t_{\rm p} \geq t_{\rm p'})$ as a function of
    processing time $\tau$, evaluated using
    Eq.~\eqref{eq:eta-dna}. (b) Accuracy $\mathcal{A}$ as a function
    of processing time $\tau$ and contact duration $T$, evaluated
    using Eq.~\eqref{eq_acc:accuracy_denotation}.  (c) The maximal
    accuracy $\mathcal{A}^*$ (squares) as a function of processing
    time $\tau$, evaluated using
    Eq.~\eqref{eq_acc:accuracy_N_star}. The asymptotic behavior of
    $\mathcal{A}^*$ in the $\tau \rightarrow \infty$ limit is shown by
    the dashed curve, evaluated using
    Eq.~\eqref{eq_acc:accuracy_N_star_asymptotic}.  (d) Total duration
    (MFPT $\mathbb{E}[t]$) of the DNA replication process as a
    function of inaccuracy $\mathbb{P}(t_{\rm p} \geq t_{\rm p'})$ is
    evaluated using Eq.~\eqref{eq:mfpt-dna} and is shown in blue. The
    optimal contact duration $T^*$ in the TCR recognition scenario as
    a function of inaccuracy $1 - \mathcal{A}^*$ (red) is evaluated
    using Eq.~\eqref{eq_acc:accuracy_N_star} and the definition $N=k_1
    T$.  In (a-d), we set $k_{1}=q_{1}=0.1$, $k_{-1}=1$, and
    $q_{-1}=2$.}
  \label{fig:fpt_summary}
\end{figure}

KPR in the DNA replication scenario relies on a comparison between the
FPT to incorporate the correct nucleotide $t_{\rm p}$ and the FPT
$t_{\rm p}'$ of incorporating an incorrect one, while KPR in the
FPT-based TCR activation scenario requires comparison of the
processing time $\tau$ with the cell-cell contact time $T$.  As shown
in Eq.~\eqref{eq:eta-dna}, the error probability $\mathbb{P}(t_{\rm p}
\geq t_{\rm p'})$ of incorporating one incorrect nucleotide scales
exponentially with respect to the processing time $\tau$ in the $\tau
\rightarrow \infty$ limit
\begin{equation}\label{eq:eta-dna-limit}
  \mathbb{P}(t_{\rm p} \geq t_{\rm p'}) \approx  \frac{q_1}{k_1}e^{-(q_{-1} - k_{-1})\tau}.
\end{equation}
Consequently, a longer processing time $\tau$ can exponentially reduce
the error probability, as shown in Fig.~\ref{fig:fpt_summary}(a). Note
that the error probability depends linearly on the binding rates
$q_{1}/k_{1}$ but exponentially on the unbinding rates $q_{-1},
k_{-1}$. This different dependence indicates the nonequilibrium nature
of the KPR process.

For convenience, in FPT-based scenario, we assume parameters $k_1 =
q_1 \ll k_{-1} \sim q_{-1}$, $T \gg k_1^{-1}$ and let $N = T k_1$,
which allows us to treat the recognition process as $N$ cycles of a
one-shot process. For each shot, an initial complex ${\rm ES}$ either
unbinds with a probability of $e^{-q_{-1}\tau}$ or $e^{- k_{-1}
  \tau}$, or activates. Then, $N$ independent cycles of the process
yields the geometric success probability $(1 - e^{- k_{-1} \tau})^N$.

Under the assumptions introduced previously, we can evaluate the
accuracy $\mathcal{A}$ defined by Eq.~\eqref{eq_acc:accuracy} as
\begin{equation} \label{eq_acc:accuracy_denotation}
  \begin{aligned}
  \mathcal{A}(\tau, N) &= \frac{1}{2} + \frac{\left(1 - e^{-q_{-1} \tau}\right)^N}{2}
   - \frac{\left(1 - e^{-k_{-1} \tau}\right)^N}{2} \\
   &= \frac{1}{2} + \frac{1}{2}\left(1 - e^{-q_{-1}\tau}\right)^N
   \left[1 - \left(\frac{1 - e^{-k_{-1} \tau}}{1 - e^{-q_{-1}\tau}}\right)^N\right].
  \end{aligned}
\end{equation}
The accuracy depends on both the number of binding events $N$ (hence
$T$) and processing time $\tau$, as illustrated in
Fig.~\ref{fig:fpt_summary}(b). For fixed contact duration $T$, the
accuracy first increases with $\tau$, then followed by a decrease. In
the long processing time limit $\tau \rightarrow \infty$, the T cell
does not respond to any signal, corresponding to
$\mathcal{A}=\tfrac{1}{2}$.

For fixed processing time $\tau$, there is a maximal accuracy
$\mathcal{A}^*(\tau)=\sup_N \mathcal{A}(\tau,N)$ and a corresponding
$N^*(\tau)$ such that $\mathcal{A}(\tau, N^*)= \mathcal{A}^*(\tau)$.
A straightforward calculation yields
\begin{equation} \label{eq_acc:N_star}
  \begin{aligned}
  N^*  \approx \frac{(q_{-1} - k_{-1})\tau}{1- e^{-(q_{-1}-k_{-1})\tau}} e^{k_{-1} \tau}
  \end{aligned}
\end{equation}
and 
\begin{equation}
  \begin{aligned}
  \mathcal{A}^* (\tau) = 
  \frac{1}{2}+ & \frac{1}{2}
  \left(1-e^{-q_{-1} \tau}\right)^{\frac{\left(q_{-1}-k_{-1}\right) \tau}
    {e^{-k_{-1} \tau}-e^{-q_{-1}\tau}}}  \times \left(1-e^{-\left(q_{-1}-k_{-1}\right) \tau}\right).
  \end{aligned}
  \label{eq_acc:accuracy_N_star}
\end{equation}
The asymptotic behavior of $\mathcal{A}^*$ in the $\tau \rightarrow
\infty$ limit is
\begin{equation}
    \label{eq_acc:accuracy_N_star_asymptotic}
    1-\mathcal{A}^*_{\rm asymp} (\tau) = \frac{(q_{-1} - k_{-1})\tau}{2}
    e^{-(q_{-1} - k_{-1})\tau} .
\end{equation}

As demonstrated by Eq.~\eqref{eq:eta-dna-limit},
Eq.~\eqref{eq_acc:accuracy_N_star_asymptotic}, and
Fig.~\ref{fig:fpt_summary}(a,c), for both DNA replication and TCR
recognition scenarios, the \textit{inaccuracy} or error probability
scales with $e^{- (q_{-1} - k_{-1})\tau}$. The improvement in the
accuracy comes at a cost of the increased total times spent in the
proofreading process. The mean time spent in the DNA replication
process is given by Eq.~\eqref{eq:mfpt-dna}.  The optimal contact
duration $T^*$ required for a specific $\tau$ is given by
$\frac{1}{k_{1}} N^*$, with $N^*$ given by
Eq.~\eqref{eq_acc:accuracy_N_star}.  In both cases, there is a common
$e^{k_{-1} \tau}$ factor.  Thus, for both scenarios, there is a
trade-off between the accuracy and the total time spent in the
proofreading process, as is shown in Fig.~\ref{fig:fpt_summary}(d).

In the following sections, we will focus primarily on the 
TCR recognition scenario.

\subsection*{Channel capacity and Fisher linear discriminant agree qualitatively}
We first show that the channel capacity metric and the Fisher linear
discriminant metric show qualitatively similar behavior. To obtain
statistically accurate results, we simulate the TCR recognition
process using the Gillespie algorithm and evaluate the channel
capacity and Fisher linear discriminant metric from $10^4$
realizations. The results shown in
Fig.~\ref{fig:multiround_simulation_CC_FLD_tau} and
Fig.~\ref{fig:multiround_simulation_CC_FLD_T} indicate that both
quantities increase with respect to cell-cell contact time $T$ and are
maximal at waiting time $\tau\sim 0.6/k_{-1}$, regardless of $T$.  We
now establish the channel capacity as an appropriate metric for
comparing different discrimination strategies.

\subsection*{Invariant optimal processing time of product-based strategy for different contact times}
\begin{figure}
  \centering 
  \includegraphics[width=4.6in]{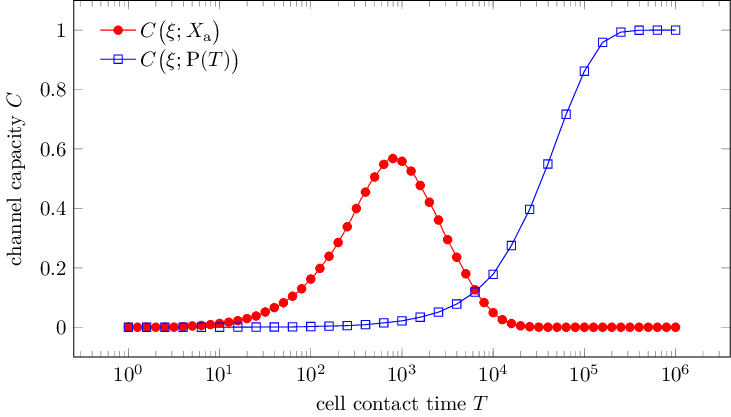}
  \vspace{4mm}
  \caption{\label{fig:channel-capacity-first-passage-time} The channel
    capacity as a function of cell-cell contact time $T$ for
    first-passage-time-based (FPT-based) signaling and product-based
    signaling. The channel capacity is evaluated between the input
    $\xi$ indicating correct (1) or incorrect (0) substrate and the
    output $X_{\rm a}$ or ${\rm P}(T)$. We assumed $k_{1}=q_1 = 0.1$,
    $k_{-1}=k_{-1}^* = 1$, $q_{-1}=q_{-1}^* = 2$, $\tau = 3$, and
    $k_{\rm p} = 0.01$ for a slow product formation rate.}
\end{figure}

We now compare the channel capacity of FPT-based discrimination to
that associated with product-based discrimination.  In
Fig.~\ref{fig:channel-capacity-first-passage-time} we plot the channel
capacity between the input $\xi$ and the outputs $X_{\rm a} =
\mathds{1}_{t_{\rm a} \leq T}$ (in the FPT-based discrimination) and
${\rm P}(T)$ (in the product-based discrimination) as a function of
cell contact time $T$ when $\tau=3$ is fixed.

\begin{figure}
  \centering
  \includegraphics[width=5.8in]{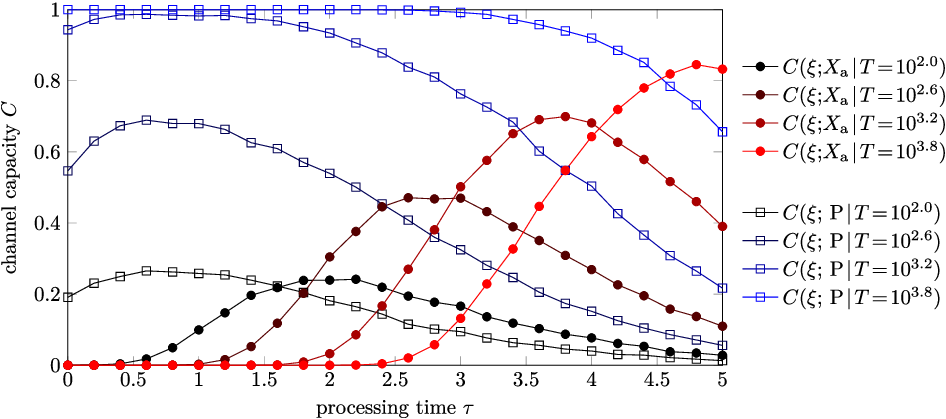}
  \vspace{4mm}
  \caption{\label{fig:channel-capacity-first-passage-time-2} The
    channel capacities of product concentration (blue squares) and
    first activation times (red dots) as a function of processing time
    $\tau$ for various cell contact times $T$. We evaluate the channel
    capacity using stochastic simulations (Gillespie algorithm) of the
    model in Eq.~\eqref{eq:multistep} with parameters $k_{1}=q_1 =
    0.1$, $k_{-1}=k_{-1}^* = 1$, $q_{-1}=q_{-1}^* = 2$, and $k_{\rm p}
    = 1$. The production rate was set higher for easier simulation of
    the product concentration.}
\end{figure}

In Fig. \ref{fig:channel-capacity-first-passage-time-2}, we plot the
channel capacity as a function of processing time $\tau$ for various
contact times $T$. As in
Fig.~\ref{fig:channel-capacity-first-passage-time}, we channel
capacities associated with both FPT-based and product-based
discrimination. The channel capacity of product concentration
increases monotonically with respect to cell contact time $T$ while
exhibiting a peak at $\tau_{\rm P}^* \sim 0.6/k_{-1}$, as is shown
previously. By contrast, the channel capacity under FPT-based
discrimination has an optimal processing time $\tau_{\rm a}^*$ that
increases with respect to cell contact time $T$. The optimal contact
time $T_{\rm a}^*$ also increases with respect to the processing time
$\tau$. The channel capacity-optimizing contact times $T_{\rm a}^*,
T_{\rm P}^*$, and processing times $\tau_{\rm P}^{*}, \tau_{\rm a}^*$
shown in Fig.~\ref{fig:channel-capacity-first-passage-time-3}(a,b).

There are two noteworthy observations from
Fig.~\ref{fig:channel-capacity-first-passage-time-2}.  First, in the
product-based scenario, the channel capacity for $\tau=0$ (no kinetic
proofreading limit) does not differ significantly from the optimal
channel capacity at $\tau^*_{\rm P}$.  This observation together with
the $T$-independent optimal processing time $\tau^*_{\rm P}$ reflects
the conclusion by Kirby et al. that KPR is ineffective due to noise
\cite{Kirby2023may}.  Second, under the same total contact duration
$T$, the maximal channel capacity of the product-based strategy is
higher than that of the FPT-based strategy provided that the
production rate $k_{\rm p}$ is sufficiently large. These two
observations suggest that the product-based discrimination is a
superior strategy compared to the FPT-based one. In order to provide
mechanistic insight into the difference between these two strategies,
we now introduce a method to analytically characterize the
product-based strategy.

\subsection*{Decomposition of the product-based
  strategy}
\begin{figure}[htbp]
    \centering
    \includegraphics[width=4.8in]{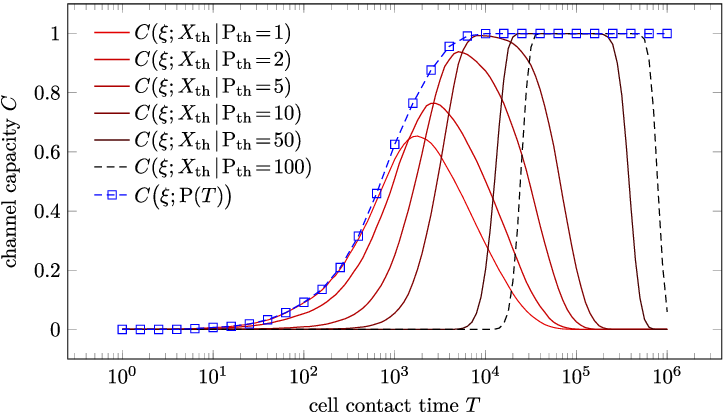}
    \vspace{4mm}
    \caption{\label{fig:channel-capacity-product-threshold} The
      channel capacity between the input $\xi$ and the output ${\rm
        P}(T)$ or $X_{\rm th}$ as a function of cell-cell contact time
      $T$. Here, $k_{1}=q_1 = 0.1$, $k_{-1}=k_{-1}^* = 1$,
      $q_{-1}=q_{-1}^* = 2$, $\tau = 3$, and $k_{\rm p} = 1$.}
\end{figure}

\begin{figure}[htbp]
  \centering
  \includegraphics[width=4.8in]{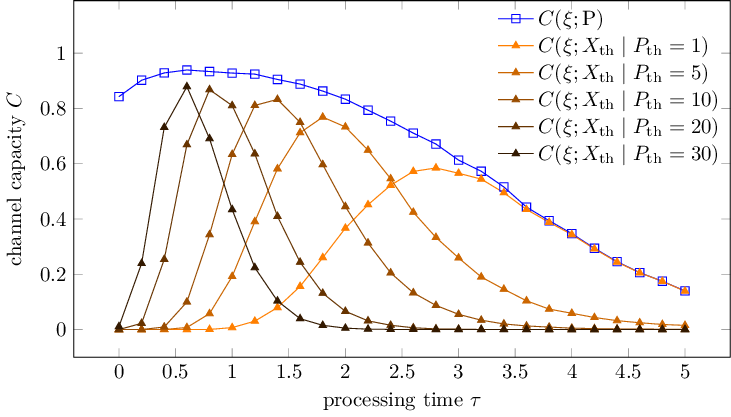}
  \vspace{4mm}
  \caption{\label{fig:channel-capacity-product-threshold-tau} The
    channel capacity between the input $\xi$ and the output $X_{\rm
      a}$, ${\rm P}(T)$, or $X_{\rm th}$ as a function of processing
    time $\tau$. We assumed $k_{1}=q_1 = 0.1$, $k_{-1}=k_{-1}^* = 1$,
    $q_{-1}=q_{-1}^* = 2$, $T = 1000$, and $k_{\rm p} = 1$. 10,000
    independent Gillespie simulations are conducted for each $\tau$.}
\end{figure}

\added{We propose decomposing the product-based strategy into a series
  of first-passage-time-based strategies using the FPT of the number
  of products ${\rm P}$ to different thresholds ${\rm P}_{\rm th}$.}
Let $t_{k}$ represent the first time the product ${\rm P}(t)$ exceeds
the threshold $k$. $X_{\rm th} = 1$ (triggering immune response) if
$t_{{\rm P}_{\rm th}} \leq T$ and $X_{\rm th} = 0$
otherwise. \added{Being a FPT-based strategy, $C(\xi; X_{\rm th})$ has
  a similar dependence on $T$ and $\tau$ to that of} $C(\xi; X_{\rm a}
)$, the channel capacity in FPT-discrimination we discussed earlier.
In order to illustrate this point, we perform Gillespie simulations of
the model in Eq.~\eqref{eq:multistep} with the same parameters as in
Fig.~\ref{fig:channel-capacity-first-passage-time-2}. The results are
shown in Fig.~\ref{fig:channel-capacity-product-threshold} and
Fig.~\ref{fig:channel-capacity-product-threshold-tau}.

Since $X_{\rm th}$ is derived from ${\rm P}(T)$, $C(\xi; {\rm P}(T))$
serves as an upper bound of $C(\xi; X_{\rm th})$ for various
thresholds $P_{\rm th}$, \added{as is illustrated in
  Fig.~\ref{fig:channel-capacity-product-threshold} and
  Fig.~\ref{fig:channel-capacity-product-threshold-tau}}. No single
threshold can reach the information upper bound $C(\xi; {\rm P}(T))$
at any $T$. A small threshold ${\rm P}_{\rm th}$ can approach $C(\xi;
{\rm P}(T))$ for small $T$ and large $\tau$, while a large threshold
${\rm P}_{\rm th}$ can approach the information upper bound $C(\xi;
{\rm P}(T))$ for large $T$ and small $\tau$.  \added{We thus introduce
  the approximation}

\begin{equation}
  \widehat{C}(\xi; {\rm P}(T)) \coloneqq \max_{{\rm P}_{\rm th}} C(\xi; X_{\rm th}) 
  \approx C(\xi; {\rm P}(T))
  \label{eq:approximation}
\end{equation}
for $C(\xi; {\rm P}(T))$.

\subsection*{Mathematical analysis of the effects $\tau$ in the product-counting strategy}
\added{Having established the approximation to the channel capacity of
  the product-based strategy in Eq.~\eqref{eq:approximation},} we
further analyze its dependence on the processing time $\tau$ under a
fixed cell-cell contact time $T$.
%
%
%
\added{We use the equivalence between channel capacity and accuracy in
  the high accuracy limit, as shown in
  Eq.~\eqref{eq:information_accuracy} to evaluate $C(\xi; {\rm
    P}(T))$. We first introduce a Gaussian distribution approximation
  to the original distribution of ${\rm P}(T)$ with matched mean and
  variance}
%
%
\begin{equation}
\begin{aligned}
    \mathbb{E}[{\rm P}(T) \mid \xi=1] &
    \approx\operatorname{var}[{\rm P}(T) \mid \xi=1] 
    \approx k_{\rm p} T K e^{-k_{-1} \tau}, \\
    K &  \equiv \frac{k_1}{k_1 + k_{-1}}.
\end{aligned}
\end{equation}
The $\xi=0$ case is similar. The above steady-state approximation is
justified in Appendix~\ref{appendix:master-equation}.

The condition $t_{\rm P_{th}} < T$ is equivalent to ${\rm P}(T) \geq
{\rm P}_{\rm th}$. Approximating the distribution of ${\rm P}(T)$
by a normal distribution with mean $\mathbb{E}\left[{\rm P}(T)\right]$
and variance $\operatorname{var}\left[{\rm P}(T)\right]$, we can
estimate the conditional probabilities of $X_{\rm th}=
\mathds{1}_{{\rm P}(T) > {\rm P}_{\rm th}}$ given $\xi$ by the
integral
\begin{equation}\label{eq:specificity-sensitivity-in}
  \begin{aligned}
   \text{sensitivity} & = \mathbb{P}(X_{\text{th}}\!=1\,\vert\, \xi =1) \\
\: & = \int^{\infty}_{
        \frac{{\rm P}_{\rm th} - k_{\rm p} T K e^{-k_{-1} \tau}}
            {\sqrt{k_{\rm p} T K e^{-k_{-1} \tau}}}}\! e^{- x^2/2}\frac{\mathrm{d}x}{\sqrt{2\pi}}, \\[5pt]
    \text{specificity} & = \mathbb{P}(X_{\text{th}}\!= 0\,\vert\, \xi = 0) \\
\: & = \int^{\frac{{\rm P}_{\rm th} - k_{\rm p} T Q e^{-q_{-1} \tau}}
        {\sqrt{k_{\rm p} T Q e^{-q_{-1}\tau}}}}_{- \infty }\!e^{- x^2/2}\frac{\mathrm{d}x}{\sqrt{2\pi}}.
      \end{aligned}
\end{equation}
Under the assumption that $\mathbb{E}[{\rm P}(T) \mid \xi=1] > {\rm
  P}_{\rm th}$ and $\mathbb{E}[{\rm P}(T) \mid \xi=0] < {\rm P}_{\rm
  th}$, we can rewrite Eq.~\eqref{eq:specificity-sensitivity-in} as
\begin{equation}
  \label{eq:specificity-sensitivity-erf}
  \begin{aligned}
    \mathbb{P}(X_{\text{th}}\!=1\,\vert\, \xi =1) &=
        \frac{1}{2} + \frac{1}{2}\operatorname{erf} \left( \frac{k_{\rm p} T K e^{-k_{-1} \tau}-{\rm P}_{\rm th}}
        {\sqrt{2k_{\rm p} T K e^{-k_{-1} \tau}}} \right), \\
    \mathbb{P}(X_{\text{th}}\! = 0\,\vert\, \xi = 0) &=
        \frac{1}{2} + \frac{1}{2}\operatorname{erf} \left( \frac{{\rm P}_{\rm th}-k_{\rm p} T Q e^{-q_{-1} \tau}}
        {\sqrt{2k_{\rm p} T Q e^{-q_{-1}\tau}}} \right).
  \end{aligned}
\end{equation}

In the high accuracy limit, the mutual information $C(\xi; X_{\rm
  th})$ between binary uniform input and binary output is approximated
by the accuracy $\mathcal{A}$ as in
Eq.~\eqref{eq:information_accuracy}. We can then obtain the optimal
threshold ${\rm P}_{\rm th}^*$ that maximizes the accuracy
$\mathcal{A}_{\rm th} =\mathbb{P}(X_{\text{th}}=1 \mid \xi =1) +
\mathbb{P}(X_{\text{th}} = 0 \mid \xi = 0)$ for given $T$ and $\tau$
to be
\begin{equation}
  {\rm P}_{\rm th}^* = k_{\rm p} T \sqrt{K e^{-k_{-1} \tau}
    Q e^{-q_{-1}\tau} } \big[1 + O(T^{-1})\big].
  \label{eq:pth_approx}
\end{equation}
The exact optimal threshold ${\rm P}_{\rm th}^*$ can be analytically
solved, but we keep the above asymptotic form for simplicity when $T
\rightarrow \infty$.
The maximal accuracy $\mathcal{A}_{\rm th}(\tau, {\rm P}_{\rm th}={\rm
  P}_{\rm th}^*)$ is then used to approximate the maximal channel
capacity $\max_{{\rm P}_{\rm th}}C(\xi; X_{\rm th} \mid {\rm P}_{\rm
  th})$, which is an approximation of the channel capacity $C(\xi;
{\rm P}(T))$ of the product-based discrimination.

\begin{figure}
    \centering 
    \includegraphics[width=4.8in]{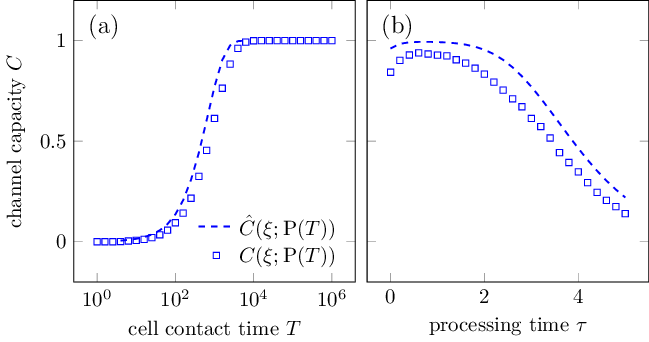}
    \vspace{4mm}
    \caption{\label{fig:cc-estimates} Comparison between the simulated
      channel capacity $C(\xi; {\rm P}(T))$ and the corresponding
      estimate using Eq.~\eqref{eq:channel_capacity_approx}. (a)
      $C(\xi; {\rm P}(T))$ as a function of cell-cell contact time
      $T$; (b) $C(\xi; {\rm P}(T))$ as a function of processing time
      $\tau$. Here, we took $k_{1}=q_1 = 0.1$, $k_{-1}=k_{-1}^* = 1$,
      $q_{-1}=q_{-1}^* = 2$, and $k_{\rm p} = 1$. $\tau=3$ in (a) and
      $T=1000$ in (b).}
\end{figure}

Substituting the zero-th order term of Eq.~\eqref{eq:pth_approx} into
Eq.~\eqref{eq:specificity-sensitivity-erf}, we find
\begin{equation}
  \begin{aligned}
  \hat{C}(\xi; & {\rm P}(T) \mid \tau) \coloneqq 
  \mathcal{A}_{\rm th}(\tau, {\rm P}_{\rm th}={\rm P}_{\rm th}^*)\\
  =  & \frac{1}{2} + \frac{1}{2} \operatorname{erf} 
  \Big(\sqrt{\tfrac{k_{\rm p}T K e^{- k_{-1} \tau}}{2}}
  - \sqrt{\tfrac{k_{\rm p}T Q e^{- q_{-1} \tau}}{2}}
  \Big)
  \end{aligned}
  \label{eq:channel_capacity_approx}
\end{equation}
In Fig.~\ref{fig:cc-estimates}, note that
Eq.~\eqref{eq:channel_capacity_approx} generally matches the
simulation results well. Slightly higher values of
Eq.~\eqref{eq:channel_capacity_approx} arise from the Gaussian
approximation used to map discrete output to continuous output.

The choice of optimal processing time $\hat{\tau}^*_{\rm P}$ that
maximizes the accuracy in Eq.~\eqref{eq:channel_capacity_approx} is
independent of $T$ and is given by
\begin{equation}
\begin{aligned}
  \hat{\tau}^*_{\rm P} &= \frac{2}{q_{-1} - k_{-1}}
\log\Big(\sqrt{\frac{Q}{K}} \frac{q_{-1}}{k_{-1}}\Big).
\end{aligned}
\label{eq:tau_star}
\end{equation}
\added{Note that the optimal processing time $\hat{\tau}^*_{\rm P}$ is
  obtained by taking ${\rm P}_{\rm th}={\rm P}^*_{\rm th}(T)$ and is
  independent of the cell-cell contact time $T$. Under the parameter
  settings where $k_1=q_1=0.1$, $k_{-1}=1$, and $q_{-1}=2$, we find
  that $\hat{\tau}^*_{\rm P}$ is approximately $0.6/k_{-1}$, which is
  consistent with the optimal processing time $\tau_{\rm P}^*$
  obtained from the simulation results, as illustrated in
  Fig.~\ref{fig:channel-capacity-first-passage-time-3}.}

\subsection*{The dynamical threshold and random cell-cell contact time}
In the previous sections, we have assumed that the cell-cell contact
time $T$ is deterministic.  In reality, the cell-cell contact time $T$
is random and vary from cell to cell
\cite{Miller2004January,Bousso2003May,Henrickson2008January}. To
evaluate the effects of a random cell-cell contact time, we consider a
simple model where the cell-cell contact time $T$ is uniformly
distributed in the interval $[0, T_{\rm max}]$, where $T_{\rm max}$ is
the maximal cell-cell contact time.

\begin{figure}[htbp]
  \centering 
  \includegraphics[width=5.2in]{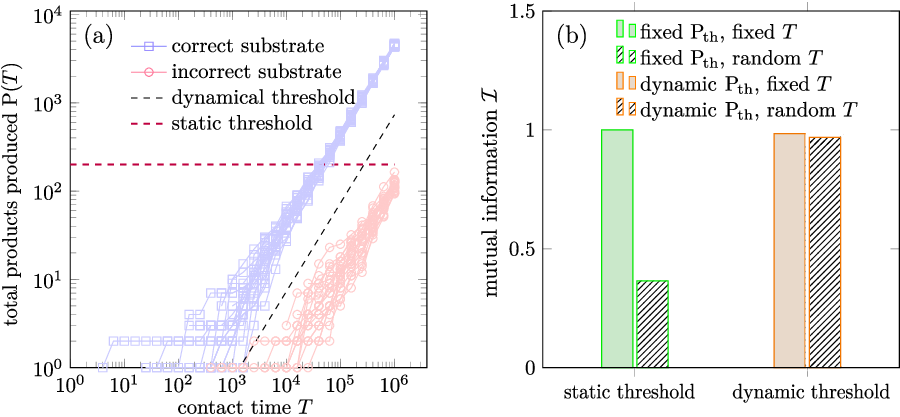}
  \vspace{4mm}
\caption{The dynamical threshold and random contact time. (a)
  Illustration of the dynamical threshold ${\rm P}_{\rm th}^*$ as a
  function of the time $t$ since initial contact. The blue
  trajectories represent the number of products ${\rm P}$ with correct
  substrates.  The red trajectories represent that of incorrect
  substrates. (b) A dynamical-threshold-based discrimination strategy
  maintains a high channel capacity when the total contact time $T$ is
  uniformly distributed between $0$ and $T_{\rm max}$. Filled bars
  represent the mutual information between input $\xi$ and output
  $X_{\rm th}$ with a fixed contact time $T$ and patterned bars
  represent the mutual information with a uniformly distributed
  contact time $T$ between $0$ and $T_{\rm max}$. The green bars
  indicate the maximal mutual information over all possible contact
  times $T \leq T_{\rm max}$ and all possible static thresholds ${\rm
    P}_{\rm th}$. The input $\xi$ is assumed to be uniformly
  distributed on $\left\{ 0,1 \right\}$.  We assumed $k_{1}=q_1 =
  0.1$, $k_{-1}=k_{-1}^* = 1$, $q_{-1}=q_{-1}^* = 2$, $\tau = 3$, and
  $k_{\rm p} = 1$. $T_{\rm max}$ is set to $10^6$. We additionally
  mandate that when the dynamical threshold ${\rm P}_{\rm th}^*(t)$ is
  smaller than 10 products, no response is initiated to filter out the
  initial noise.}
\label{fig:dynamical_threshold}
\end{figure}

In the previous section, we conclude that the $T$-independent optimal
processing time $\tau_{\rm P}^*$ is a result of fixing the threshold
${\rm P}_{\rm th}$ to ${\rm P}_{\rm th}^*(T)$ given by
Eq.~\eqref{eq:pth_approx}.  In the case of a random cell-cell contact
time $T$, choosing a universally optimal threshold ${\rm P}_{\rm
  th}^*$ is difficult. We can, however, choose a dynamical threshold
${\rm P}_{\rm th}^*(t)$ that increases with the time $t$ passed since
the initial contact to maximize the channel capacity, where ${\rm
  P}_{\rm th}^*$ is still given by Eq.~\eqref{eq:pth_approx}, with $T$
replaced by $t$. A comparison of a dynamical threshold ${\rm P}_{\rm
  th}^*(t)$ and a static threshold ${\rm P}_{\rm th}$ is shown in
Fig.~\ref{fig:dynamical_threshold}(a).

The numerical results of the maximal mutual information between the
input and output $X_{\rm th}$ under different contact times $T$ and
different thresholds ${\rm P}_{\rm th}$ are shown in
Fig.~\ref{fig:dynamical_threshold}(b). In the case of a fixed contact
time $T$, the maximal mutual information of both static and dynamical
thresholds is close to 1, indicating perfect discrimination.  In the
case of a uniformly distributed contact time $T$ between $0$ and
$T_{\rm max}$, the maximal mutual information of the dynamical
threshold is close to 1, while that of the static threshold is close
to 0.4, indicating poor discrimination.

The experiments in Fig.~\ref{fig:dynamical_threshold} suggest that the
dynamical threshold ${\rm P}_{\rm th}^*(t)$ maintains a high channel
capacity when the total contact time $T$ is random.

\subsection*{A nested single-binding KPR scheme}
We have provided a mechanistic explanation for the different
qualitative behaviors of the channel capacity between
first-passage-time-based discrimination and product-based
discrimination, in particular, why the optimal processing time
  $\tau_{\rm P}^*$ is independent of the cell-cell contact time $T$ in
  the product-based discrimination.

It is also of interest to understand the high channel capacity in the
product-based discrimination problem that arises even in the $\tau \to
0$ limit, as shown in the blue curve of $C(\xi; {\rm P})$ in
Fig.~\ref{fig:channel-capacity-product-threshold-tau}.  To provide
simple mechanistic intuition, note that production and accumulation of
products can be \replaced{comparable to}{realized by} a sequence of
phosphorylation events in the traditional KPR scheme. The cell-cell
contact time $T$ sets the rate of the competing unbinding event:
\begin{equation}\label{eq:nested-KPR-scheme}
    {\rm APC} + {\rm T} \stackrel{\tilde k_{\pm 1}}{\rightleftharpoons} 
    \!\text{APC-T}^{(0)}\!\stackrel{\tilde k_{\rm f}}{\rightarrow} 
    \!\text{APC-T}^{(1)}\!\!\rightarrow \cdots \rightarrow \!\text{APC-T}^{({\rm P_{\rm th}})}.
\end{equation}
\added{Here, ATC represent the antigen-presenting cell (APC) and the T
  cell is denoted by T. The superscript $(i)$ represents the number of
  products generated by the T cell. The T cell initiates responses
  when the number of products exceeds a threshold ${\rm P_{\rm th}}$.
  The overall scheme is comparable to the traditional KPR scheme shown
  in Fig.~\ref{eq:reaction}(a).}
The difference between the traditional KPR scheme and the
product-based strategy shown above is that the latter assumes a
different phosphorylation rate $\tilde k_{\rm f} \equiv k_{\rm p} K
e^{-k_{-1} \tau}$ or $k_{\rm p} Q e^{-q_{-1}\tau}$ for correct and
incorrect substrates, while the unbinding event has a common timescale
$T$ set by the cell-cell contact time.

We have assumed that the contact time $T$ is approximately constant
given that unbinding of the TCR and the APC requires multiple steps
involving collective effects of adhesion membrane proteins.  Also note
that the activation time of the TCR signaling process, given by the
FPT $\tau_{\rm P_{\rm th}}$ to generate a number of products ${\rm P}$
beyond the threshold ${\rm P}_{\rm th}$ is also approximately a
constant when ${\rm P_{\rm th}}$ is sufficiently large.
Consequently, the competition between two processes with almost fixed
timescales allows highly informative outputs compared to the
traditional KPR scheme and Michaelis-Menten kinetics, as indicated in
Eq.~\eqref{eq:nested-KPR-schematic}, where $\rightsquigarrow$ and
$\leftsquigarrow$ represent a deterministic waiting time $\tau$.

\begin{equation}
    \begin{aligned}
   \text{ Michaelis-Menten: } & \rm E+S \longleftarrow E S \longrightarrow E^* S, \\
        \text { KPR: } & \rm  E+S \longleftarrow E S \rightsquigarrow E^* S, \\
        \text { nested KPR: } & \rm E+S \leftsquigarrow E S \rightsquigarrow E^* S.
    \end{aligned}
    \label{eq:nested-KPR-schematic}
\end{equation}

To further illustrate the role of a deterministic waiting time
in the accuracy of the above process, we compare it with
an exponential waiting time, \textit{i.e.}, standard Michaelis-Menten (MM)
kinetics in Appendix~\ref{analysis_of_information}. As for a comparison 
between the traditional KPR scheme and the nested KPR scheme, we
note that in the limit of deterministic waiting time, the nested KPR 
can achieve exact discrimination.

\section*{Discussion}
In this paper, we have considered kinetic proofreading schemes for two
classes of biological processes, DNA replication process and T cell
signaling.  Overall, our analysis indicates that how the output of a
kinetic proofreading process is used to make a decision is crucial to
the performance of KPR.  We have shown that in the case of DNA
replication, the specificity of the process is always exponentially
dependent on the processing time $\tau$ and increasing specificity
comes at the cost of replication speed.

In the case of TCR signaling, the trade-off between specificity and
sensitivity can be mitigated by increasing the number of allowed
failed attempts $N$ which is proportional to the cell-cell contact
time $T$. The overall accuracy $\mathcal{A}$ of the signaling process
is still exponentially dependent on the processing time $\tau$, as
illustrated by Eq.~\eqref{eq_acc:accuracy_N_star_asymptotic}.  For
longer processing time $\tau$, the higher accuracy can only be
achieved at the cost of the signaling speed by exponentially
increasing the cell-cell contact time $T$, or equivalently, the number
of allowed failed attempts $N$ to the optimal value $N^*$ indicated by
Eq.~\eqref{eq_acc:N_star}.

A variant of the FPT-based strategy, extracting the extreme FPT in the
presence of multiple substrates (antigens), has also been proposed
\cite{Morgan2023aug}. The main goal in \cite{Morgan2023aug} is to
determine sensitivity and specificity of TCR recognition of foreign
antigens when also exposed to a sea of self-antigens. The
self-antigens in \cite{Morgan2023aug} are assumed to bind much more
weakly to the TCR than the foreign antigens. While we are primarily
interested in discrimination between correct and incorrect substrates
with similar binding affinities. This is a typical situation as T
cells need to identify cancer cells that present mutated self antigens
on their surface. In this case, the affinity of the corresponding TCR
to the self antigen is expected to be similar to the affinity to the
foreign antigen. The densities of the self and foreign antigens are
also expected to be similarly low.

The amount of product produced can also be used to distinguish correct
and incorrect substrates. To compare the performance of product-based
discrimination to that of FPT-based discrimination, we introduced a
channel capacity between the input $\xi \in \{0,1\}$ (incorrect,
correct substrate) and the output $X$, which can be either $X_{\rm
  a}=\mathds{1}_{t_{\rm a} \leq T}$ or the number of products
generated ${\rm P}(T)$. We established the connection between the
channel capacity and the accuracy of the discrimination problem in the
high accuracy limit in Eq.~\eqref{eq:information_accuracy}. Thus, the
dependence of the first-passage-time-based discrimination on the
processing time and the cell-cell contact time can be shown to have a
single peak, corresponding to the optimal processing time $\tau_{\rm
  a}^*$ and the optimal cell-cell contact time $T_{\rm a}^*$,
respectively, as illustrated in
Figs.~\ref{fig:channel-capacity-first-passage-time}--\ref{fig:channel-capacity-first-passage-time-2}.

By contrast, the product-counting discrimination problem has a
distinct monotonic increase of the channel capacity with respect to
the cell-cell contact time $T$ and the optimal processing time
$\tau_{\rm P}^*$ is independent of $T$. We find analytic
approximations of the channel capacity by decomposing the
product-based discrimination problem into a series of
first-passage-time-based discrimination problems with different
thresholds. This observation allows us to analytically approximate
$\tau_{\rm P}^*$ by Eq.~\eqref{eq:tau_star}. The approximation in turn
reveals that the peculiar dependence of the channel capacity on the
processing time and the cell-cell contact time arises from choosing an
optimal threshold ${\rm P}_{\rm th}^*$ for distinguishing correct
substrates from incorrect ones.



Our TCR signaling model is a simplified scenario that assumes a
deterministic cell-cell contact time $T$ and a deterministic
processing time $\tau$. However, we conducted additional simulations
that relaxed the deterministic processing time assumption by explicit
modeling of the irreversible phosphorylation process in
Appendix~\ref{appendix:multistep_KPR}. The results show qualitatively
similar behavior as the simplified model. Additionally, we fixed the
mean processing time and varied the number of processing steps $m$ to
explore the effect of the number of processing steps in KPR on the
channel capacity. The results suggest that larger $m$ increases the
channel capacity of the FPT-based discrimination scheme, while the
channel capacity of the product-based discrimination remains
effectively unchanged, as shown in
Fig.~\ref{fig:multiround_multistep_simulation_CC_FLD_T_1000}.

\added{A random cell-cell contact time $T$ may also impair the
  performance of the T cell in distinguishing correct ligands from
  incorrect ones. The dynamical threshold ${\rm P}_{\rm th}$ may
  rescue this impaired performance by allowing the T cell to
  effectively adjust the threshold in time. This rescue is illustrated
  in Fig.~\ref{fig:dynamical_threshold}. However, implementing a
  dynamic threshold requires the T-cell to keep track of the duration
  since the initial contact with the APC to adjust the threshold ${\rm
    P}_{\rm th}$ accordingly.  The duration may be tracked by another
  series of similar mechanical or biochemical reactions on the
  membrane-membrane interface that are triggered by the
  membrane-membrane contact, as discussed in \cite{Fu2023}.
  Experimentally, the presence of such a dynamical threshold can be
  detected by simultaneously measuring the number of products ${\rm
    P}(T)$, the total contact time $T$ until full activation, and
  other markers indicating whether the T-cell is activated or not.}

Alternatives to the dynamical threshold strategy have been introduced
in the literature, including adaptive KPR models
\cite{AltanBonnet2005oct,Lalanne2013may,Tischer2019apr,
  Pettmann2021may,Voisinne2022aug}, force-dependent signaling
including catch bonds \cite{Liu2014apr,Sibener2018jul}, and KPR through 
spatial gradient \cite{KPR_spatial}. Our analysis
approach can also be applied to these different contexts to provide
better analytical understanding of the interplay between speed and
accuracy.

\added{The main observation that motivates the introduction of
  different strategies for discrimination is that TCR signaling is not
  an isolated process. Rather, signaling processes are an integral
  part of the cellular reaction network. It would be interesting to
  explore how the information is transferred from the TCR signaling
  process to the downstream reaction networks. Such investigations may
  provide insights into a long-standing but often overlooked question:
  what type of information is transferred from signaling process to
  the downstream pathways? Our work assumes that information is
  primarily transmitted via a binary signal that decides whether a T
  cell response is triggered, while Kirby et al. \cite{Kirby2023may}
  assumed that the information is a continuous signal that reflects
  the strength of binding affinity between ligands and receptors.}

On a theoretical level, kinetic signaling schemes represent
stochastic, biological implementations of the classical Maxwell's
demon \cite{Rex2003}, where the receptor is the demon that measures
the affinity of the ligand to the receptor and sorts the ligands
accordingly. The canonical Maxwell demon needs memory to measure both
the position and time of a particle.  In the case of a stochastic
demon and the measurement of binding affinity, a memoryless
exponential processing time seems to be able to provide a nonzero
channel capacity to distinguish the correct and incorrect
ligands. However, additional memory as provided by the nonequilibrium
kinetic proofreading process (and non-exponentially distributed
waiting time $\tau$) enhances the channel capacity
significantly. While previous literature has explored the idea of a
Maxwell's demon \cite{Manzano2021February,Cao2009April} and
energy-accuracy bounds in generalized KPR processes
\cite{Yu2022March}, the quantitative interplay between energy cost,
memory, and information processing still lacks a suitable language and
awaits future elucidation. \added{In particular, we have not
  considered energy cost which may influence the preference in cells
  for different strategies of discriminating correct substrates from
  incorrect substrates.}


\section*{Supporting information}

\paragraph*{S1 Appendix.} \textbf{Mathematical appendices and additional figures.}
\textbf{Appendix A1}: master equation for the simplified KPR
model. \textbf{Appendix A2}: analysis of information transmitted by
KPR and Michaelis-Menten schemes. \textbf{Appendix A3}: multistep
binding model. \textbf{Appendix A4}: additional figures.



\section*{Acknowledgments}
The authors thank A. Zilman and X. Guo for insightful discussions on
KPR and the manuscript.

\nolinenumbers

\bibliography{stochastic_KPR_arxiv}

\newpage

\noindent {\LARGE \textbf{S1 Supporting Information}}

\vspace*{0.15in}
\setcounter{figure}{0}
\renewcommand{\thefigure}{A\arabic{figure}}
\renewcommand{\theequation}{A\arabic{equation}}
\renewcommand{\thesection}{A\arabic{section}}

\section{Master equation for the stochastic KPR model}\label{appendix:master-equation}
Here, we consider the master equation associated with the multi-round
proofreading model described in the TCR setting of the main text and
derive numerical methods to evaluate the mean and variance of the
product ${\rm P}(T)$ produced up to time $T$.

We use $\mathbb{P}(n, {\rm E}+{\rm S}; t)$ to denote the probability
at time $t$ that the system in the ${\rm E}+{\rm S}$ state and that
exactly $n$ products exist. Similarly, we use $\mathbb{P}(n, {\rm
  E}{\rm S};t)$ and $\mathbb{P}(n, {\rm E}^{*}{\rm S};t)$ to denote
the probability of $n$ products and the system in the ${\rm E}{\rm S}$
and ${\rm E}^{*}{\rm S}$ states, respectively.  However, since our
model involves a deterministic processing time $\tau$, we introduce
the probability density function $\rho(n,a; t)$, where $a$ indicates
the age of the complex since its formation.  $\mathbb{P}(n,{\rm E}{\rm
  S};t)$ and $\mathbb{P}(n,{\rm E}^{*}{\rm S};t)$ are then given by
\begin{equation}
    \begin{aligned}
      \mathbb{P}(n,{\rm E}{\rm S};t) = & \int_0^\tau \!\rho(n,a;t) \dd a, \\
      \mathbb{P}(n,{\rm E}^{*}{\rm S};t) = & \int_\tau^\infty \!\rho(n,a;t) \dd a.
    \end{aligned}
\end{equation}

The off rates can be considered as a death rate of the age-structured
complexes ${\rm E}{\rm S}$ and ${\rm E}^*{\rm S}$, while the on rate
$k_1$ times the probability of the ${\rm E}+{\rm S}$ state is a birth
rate of ${\rm E}{\rm S}$.  Consequently, an age-structured master
equation can be written as

  \begin{equation}
    \begin{aligned}
      \frac{\mathrm{d}}{\mathrm{d} t} \mathbb{P}(n, {\rm E}+{\rm S};t) &
      =-k_1 \mathbb{P}(n, {\rm E}+{\rm S};t )
      +k_{-1}  \mathbb{P}(n, {\rm E} S;t) + k_{-1}^*
      \mathbb{P}\left(n, {\rm E}^{*}{\rm S};t\right), \\
        \rho(n, 0;t) & = k_1 \mathbb{P}(n, {\rm E}+{\rm S};t), \\
        \partial_t \rho(n,a;t) + \partial_a \rho(n,a;t) & = -k_{-1} \rho(n,a;t),
        \quad a \leq \tau, \\
        \partial_t \rho(n, a;t) + \partial_a \rho(n, a;t) & = -k_{-1}^* \rho(n, a;t)
        -k_{\rm p}\rho(n, a;t)+k_{\rm p} \rho\left(n-1, a;t\right), \quad a > \tau.  \\
        \end{aligned}
\end{equation}

The average of the number of products ${\rm P}$ at time $T$ is defined by
\begin{equation}
    \mathbb{E}[{\rm P}(T)] = \langle n \rangle_T
    = \sum_{n=0}^\infty n \left[\mathbb{P}(n, {\rm E}+{\rm S};T)
      + \mathbb{P}(n, {\rm E}{\rm S};T)
    +  \mathbb{P}(n, {\rm E}^{*}{\rm S};T)\right]
\end{equation}
and satisfies the ODE
\begin{equation}
  \frac{\mathrm{d}}{\mathrm{d}t} \langle n(t) \rangle =
  k_{\rm p} \mathbb{P}({\rm E}^{*}{\rm S};t),
    \label{eq:mean}
\end{equation}
where $\mathbb{P}({\rm E}^{*}{\rm S};t)\equiv \sum_{n=0}^{\infty}
\mathbb{P}(n,{\rm E}^{*}{\rm S};t)$ is the marginal probability of the
${\rm E}^{*}{\rm S}$ state. Similarly, to find the variance of ${\rm
  P}(T)$, we first find the dynamics of $\langle n^2(t) \rangle$ as
follows
\begin{equation}
    \frac{\mathrm{d}}{\mathrm{d} t}\left\langle n^2(t)\right\rangle = 2 k_{\rm p}
  \sum_{n=0}^{\infty} n \mathbb{P}\left(n,{\rm E}^{*}{\rm S};t\right)
  +k_{\rm p} \mathbb{P}\left({\rm E}^{*}{\rm S};t\right).
    \label{eq:mean2}
\end{equation}
We denote $\sum_{n=0}^{\infty} n \mathbb{P}\left(n,{\rm E}^{*}{\rm
  S};t\right)$ by $\langle n(t)\vert {\rm E}^{*}{\rm S}
\rangle$. Similarly, we have $\mathbb{P}({\rm E}^{*}{\rm S};t) =
\sum_{n=0}^{\infty} \mathbb{P}(n, {\rm E}^{*}{\rm S};t)$ and the
like. We develop a system of equations to solve for $\mathbb{P}({\rm
  E}^{*}{\rm S};t)$:
\begin{equation}
    \begin{aligned}
      \partial_t \mathbb{P}\left({\rm E} + {\rm S};t\right) &
      =-k_1 \mathbb{P}({\rm E}+{\rm S};t)
      +k_{-1} \mathbb{P}({\rm E} S;t)+k_{-1}^* \mathbb{P}\left({\rm E}^{*}{\rm S};t\right) \\
        \partial_t \rho(a;t) + \partial_a \rho(a;t) & = -k_{-1} \rho(a;t), \quad a<\tau \\
        \partial_t \rho(a;t) + \partial_a \rho(a;t) & = -k_{-1}^* \rho(a;t), \quad a>\tau \\
        \rho(0;t) & = k_1 \mathbb{P}({\rm E}+{\rm S};t).
        \end{aligned}
        \label{eq:es-full}
\end{equation}

Since we are interested in finding $\mathbb{P}({\rm E}^{*}{\rm S};t)$
alone, we can integrate Eqs.~\eqref{eq:es-full} over age $a$ to find
the reduced system of equations
\begin{equation}
    \begin{aligned}
      \partial_t \mathbb{P}\left({\rm E} + {\rm S};t\right) &
      =-k_1 \mathbb{P}({\rm E}+{\rm S};t)+
      k_{-1} \mathbb{P}({\rm E} S;t)+k_{-1}^* \mathbb{P}\left({\rm E}^{*}{\rm S};t\right) \\
        \partial_t \mathbb{P}({\rm E}{\rm S};t) & =k_1 \mathbb{P}({\rm E}+{\rm S};t)
      - k_{-1} \mathbb{P}({\rm E}{\rm S};t) - \rho(\tau;t) \\
        \partial_t \mathbb{P}({\rm E}^{*}{\rm S};t) &= \rho(\tau;t)
      - k_{-1}^* \mathbb{P}({\rm E}^{*}{\rm S};t) 
        \end{aligned}
        \label{eq:{es-star}}
\end{equation}
Eq.~\eqref{eq:{es-star}} is not closed because of the density term $\rho(\tau;t)$,
but we can use a mean-field approximation by assuming that
the internal relaxation of $\rho(a;t)$ is very fast. Then,  we have
$\rho(a;t) = \rho(a=0;t)e^{-k_{-1}a}$, allowing us to approximate 
\begin{equation}
    \rho(\tau;t) = \frac{k_{-1}}{e^{k_{-1}\tau}-1} \mathbb{P}({\rm E}{\rm S};t).
    \label{eq:rho-tau}
\end{equation}
Substituting Eq.~\eqref{eq:rho-tau} into Eq.~\eqref{eq:{es-star}}, we
obtain a closed system of equations for $\mathbb{P}({\rm E}^{*}{\rm
  S};t)$ and $\mathbb{P}({\rm E}{\rm S};t)$.  The approximation is
valid in the sense that the steady state of the system is consistent
with the steady state of full system described by
Eq.~\eqref{eq:es-full}. In particular,
\begin{equation}
    \mathbb{P}({\rm E}^*{\rm S};t\to\infty) 
=\frac{\frac{k_1}{k_{-1}^*} e^{-k_{-1} \tau}}{\frac{k_1}{k_{-1}^*} e^{-k_{-1} \tau}
      +\frac{k_1}{k_{-1}}\left(1-e^{-k_{-1} \tau}\right)+1}.
\label{eq:steady-state}
\end{equation}

Similarly, we develop a system of equations as follows to solve for
$\langle n(t)\vert {\rm E}^{*}{\rm S} \rangle$ using the same approximation in
Eq.~\eqref{eq:rho-tau}:
\begin{equation}
    \begin{aligned}
      & \frac{\mathrm{d}}{\mathrm{d} t}\langle n(t) \vert {\rm E}+{\rm S}\rangle
      =-k_1\langle n(t)\vert {\rm E}+{\rm S}\rangle
      +k_{-1}\langle n(t) \vert {\rm E}{\rm S}\rangle+
      k_{-1}^*\left\langle n(t) \vert {\rm E}^{*}{\rm S}\right\rangle \\
        & \frac{\mathrm{d}}{\mathrm{d} t}\langle n(t) \vert {\rm E} {\rm S}\rangle
      =k_1\langle n(t) \vert {\rm E}+{\rm S}\rangle
      -k_{-1}\langle n(t) \vert {\rm E} {\rm S}\rangle
      -\frac{k_{-1}}{e^{k_{-1}\tau}-1}\langle n(t) \vert {\rm E} {\rm S}\rangle \\
      & \frac{\mathrm{d}}{\mathrm{d} t}\left\langle n(t) \vert {\rm E}^{*}{\rm S}
      \right\rangle
      =\frac{k_{-1}}{e^{k_{-1} \tau}-1}\langle n(t) \vert {\rm E} {\rm S}\rangle -
      k_{-1}^*\left\langle n(t) \vert {\rm E}^{*}{\rm S}\right\rangle
      +k_{\rm p} \mathbb{P}\left({\rm E}^{*}{\rm S};t\right)
    \end{aligned}
    \label{eq:n_e_s}
\end{equation}
The initial conditions are given by $\langle n\vert Y \rangle_0 = 0$
for $Y = {\rm E}+{\rm S}, {\rm E}{\rm S}, {\rm E}^{*}{\rm S}$. The
analytic solutions to Eq.~\eqref{eq:n_e_s} are not tractable but
formal solutions can be obtained by Laplace transform.  Uisng the
approximation $\mathbb{P}({\rm E}^{*}{\rm S};t) \equiv \mathbb{P}({\rm
  E}^{*}{\rm S};t\to\infty)$ given in Eq.~\eqref{eq:steady-state}, the
Laplace transform of Eqs.~\eqref{eq:n_e_s} is given by
\begin{equation}
    \begin{aligned}
      s \mathcal{L}_{n, {\rm E}+{\rm S}} = & -k_1 \mathcal{L}_{n, {\rm E}+{\rm S}}
      +k_{-1} \mathcal{L}_{n, {\rm E}{\rm S}}
      +k_{-1} \mathcal{L}_{n, {\rm E}^{*}{\rm S}} \\
      s \mathcal{L}_{n, {\rm E}{\rm S}} = & k_1 \mathcal{L}_{n, {\rm E}+{\rm S}}
      -\left(k_{-1}+\tilde{k}_{-1}\right)
      \mathcal{L}_{n, {\rm E}{\rm S}} \\
        s \mathcal{L}_{n, {\rm E}^{*}{\rm S}} = & \tilde{k}_{-1} \mathcal{L}_{n, {\rm E}{\rm S}}
      -k_{-1}^* \mathcal{L}_{n, {\rm E}^{*}{\rm S}}
      +\frac{k_{\rm p} \mathbb{P}_{t\to \infty}\left({\rm E}^{*}{\rm S}\right)}{s}
    \end{aligned}
\end{equation}
where $\mathcal{L}_{n,Y}\equiv \int_{0}^{\infty} \langle n \vert
Y\rangle_t e^{-st}\dd t$ and $\tilde k_{-1} \equiv
\frac{k_{-1}}{e^{k_{-1} \tau}-1}$.  The solution can be expressed as
\begin{equation}
    \mathcal{L}_{n, {\rm E}^{*}{\rm S}}=\frac{k_{\rm p}
      \mathbb{P}\left({\rm E}^{*}{\rm S};t\to\infty\right)}{s}
    \left[s+k_{-1}^*-\frac{\tilde{k}_{-1}}{\frac{\left(s+k_1\right)\left(s+k_1+\tilde{k}_{-1}\right)}{k_1
          k_{-1}^*} -\frac{k_{-1}}{k_{-1}^*}}\right]^{-1}.
\end{equation}

\subsection{$k_{\rm p} \rightarrow 0$ limit}
Taking the $k_{\rm p} \rightarrow 0$ limit, we find
\begin{equation}
    \mathbb{P}({\rm E}^{*}{\rm S};t\to\infty)
  = \frac{\frac{k_1}{k_{-1}^*} e^{-k_{-1} \tau}}
    {1 + \frac{k_1}{k_{-1}} \left(1 - e^{-k_{-1} \tau}\right) 
       + \frac{k_1}{k_{-1}^*} e^{-k_{-1} \tau}}, \,\,\, \forall  t = O\left(\frac{1}{k_{\rm p}}\right).
       \end{equation}
Equation~\eqref{eq:mean} gives 
\begin{equation}
    \langle n(t) \rangle = t \mathbb{P}({\rm E}^*{\rm S};t\to\infty).
    \label{eq:mean-approx}
\end{equation}
Additionally, separation of time scales allows us to approximate the solution to 
Eq.~\eqref{eq:n_e_s} with 
\begin{equation}
    \langle n(t)\vert Y \rangle = \langle n(t) \rangle \mathbb{P}(Y;t)
  = \mathbb{P}(Y;t\to\infty) \mathbb{P}({\rm E}^{*}{\rm S};t\to\infty)t,
  \quad Y \equiv {\rm E}+{\rm S}, {\rm E}{\rm S}, {\rm E}^{*}{\rm S}.
\end{equation}

In particular, Equation~\eqref{eq:mean2} becomes
\begin{equation}
\begin{aligned}
    \frac{\mathrm{d}}{\mathrm{d} t}\langle n^2(t) \rangle
    & = 2 k_{\rm p} \langle n(t)\vert {\rm E}^{*}{\rm S} \rangle
    + k_{\rm p} \mathbb{P}({\rm E}^{*}{\rm S};t) \\
    & = 2 k_{\rm p} \big[\mathbb{P}({\rm E}^{*}{\rm S};t\to\infty)\big]^{2} t
    + k_{\rm p} \mathbb{P}({\rm E}^{*}{\rm S};t).
%
%
\end{aligned}
\end{equation}
Since $\langle n^{2}(t=0)\rangle = 0$, we can integrate the above equation to
find
\begin{equation}
  \langle n^2(t) \rangle  = k_{\rm p}\mathbb{P}({\rm E}^{*}{\rm S};t) t \Big[
  1+ \mathbb{P}({\rm E}^{*}{\rm S};t\to\infty) t\Big]
  \end{equation}
In other words, the variance of ${\rm P}$ at time $t$ is given by
\begin{equation}
  \mathrm{Var}[{\rm P}(t)] = \mathbb{P}({\rm E}^{*}{\rm S};t\to\infty)t
  = \mathbb{E}[{\rm P}(t)].
    \label{eq:variance-approx}
\end{equation}
One can further verify that the distribution of ${\rm P}(t)$ should be 
Poisson in the limit $k_{\rm p} \rightarrow 0$.

\subsection{Limit of discrete processes}
As we have discussed, our deterministic processing time $\tau$ can be 
considered as a limit of a discrete $m$-step KPR process with $k_{\rm
  f} = \frac{m}{\tau}$, $N \rightarrow \infty$.  
provided expressions for the mean and variance of the product ${\rm
  P}$ at time $T$ when $k_{-1}=k_{-1}^*$:
\begin{equation}
    \begin{aligned}
            \langle n(t)\rangle_m= & k_{\rm p}
      t\left(\frac{k_{f}}{k_{f}+k_{-1}}\right)^m \frac{k_1}{k_1 + k_{-1}}, \\
      \sigma_m^2(t)= & \langle n(t)\rangle_m
      \left(1+2 \tfrac{k_{\rm p}}{k_{\mathrm{off}}}\right) \\
            \:   & \quad  +\frac{2 t k_{\rm p}^2 (k_1/k_{-1})^2(2+k_1/k_{-1}+
              \tfrac{k_{\rm off}}{k_{\rm f}}(2+m+k_1/k_{-1}+m k_1/k_{-1}))}{(1+k_{-1}/k_{\rm f})^{2 m+1}(1+k_1/k_{-1})^3 k_{\mathrm{off}}}.
    \end{aligned}
\end{equation}

In the limit of $m \rightarrow \infty$, we have 
\begin{equation}
    \begin{aligned}\label{eq:mean-variance-discrete}
      \langle n(T) \rangle_m & \rightarrow
      \frac{ k_{\rm p} T e^{-k_{\rm f}}k_1}{k_1 + k_{-1}}, \\
      \sigma_m^2(T) & \rightarrow \langle n(T)\rangle_{m}
      \left(1+\frac{2 k_{\rm p}}{k_{-1}}+2
        \frac{\tau}{T} \langle n(T) \rangle_{m}\right).
    \end{aligned}
\end{equation}

When $\tau \ll T$ and $k_{\rm p} \rightarrow 0$, we have recovered the
results in Eq.~\eqref{eq:mean-approx} and
Eq.~\eqref{eq:variance-approx}.

\section{Information transmitted by KPR and Michaelis-Mentens schemes}
\label{analysis_of_information}

In order to explain why the simplified KPR scheme performs better than
the MM scheme, which is different only through a different choice of
proofreading/processing time $\tau$, we consider the DNA replication
scenario.

We will provide a mathematical analysis as well as physical intuition
from an information-theoretic perspective.  To be specific, we
consider the following two schemes:
\begin{align}
   \text{KPR:}&~ \mathrm{E} + \mathrm{S} \leftrightarrow \mathrm{ES} \stackrel{\tau^{-1}}{\rightsquigarrow} \mathrm{E}+\mathrm{P}, 
     \label{eq:scheme1} \\
    \text{MM:}&~ \mathrm{E}+\mathrm{S} \leftrightarrow \mathrm{ES} \stackrel{k_{\rm f}}{\rightarrow} \mathrm{E}+\mathrm{P}. 
    \label{eq:scheme2}
\end{align}
\subparagraph*{Mathematical analysis.}
In the scenario of DNA replication, we note that
Scheme~\eqref{eq:scheme1} yields an error probability of 
\begin{equation}
    P_{\mathrm{e},{\rm KPR}}=
    \frac{q_1e^{(k_{-1}-q_{-1}) \tau}}{k_1+q_1 e^{(k_{-1}-q_{-1}) \tau}} \rightarrow 0,\,
    \text{ as } \tau \rightarrow \infty.
\end{equation}
On the other hand, Scheme~\eqref{eq:scheme2} yields an error probability of 
\begin{equation}
    P_{\mathrm{e},{\rm MM}}=
    \frac{q_1 k_{\rm f}}{(k_1+q_1)(q_{-1}+k_{\rm f})-k_1 k_{-1}
      \frac{q_{-1}+k_{\rm f}}{k_{-1}+k_{\rm f}}-q_1 q_{-1}}.
\end{equation}

By investigating $\frac{\mathrm{d}P_{\mathrm{e}, {\rm
      MM}}^{-1}}{\mathrm{d}k_{\rm f}}$, we find that $P_{\mathrm{e},{\rm
    MM}}$ is monotonically increasing with respect to $k_{\rm f}$, with the
maximum $P_{\mathrm{e},{\rm MM}}(k_{\rm f}= \infty) = \frac{q_1}{k_1+q_1}$
and minimum $P_{\mathrm{e}, {\rm MM}} (k_{\rm f}=0+) =
\frac{q_1}{(k_1+q_1)+\frac{k_1}{k_{-1}}(q_{-1}-k_{-1})}.$

To summarize, the error probability of Scheme \eqref{eq:scheme1}
converges to 0 at an exponentially fast rate, while the error
probability of Scheme \eqref{eq:scheme2} converges to a positive limit
at a finite rate. In the no-proofreading limit, $k_{\rm f} = \infty$ and
$\tau = 0$, both schemes yield $ \frac{q_1}{k_1 + q_1} $, which
depends only on the binding rate.

\subparagraph*{Physical intuition.}
Next, we provide an intuitive explanation of the above results from an
information-theoretic perspective. To simplify the analysis, we
consider the information yield of the two schemes \textit{per
binding-unbinding cycle}. We begin by introducing a timer $a$
that records the time spent in the bound state (${\rm ES}$).

At time $t=0$, we set the system in the bound state and set the timer
$a$ to 0.  The unbinding waiting time $\tau_{-1}$ is
exponential with rate $k_{-1}$ or $q_{-1}$, depending on the input
$\xi=1$ or $\xi=0$. At the moment of unbinding, the timer $a
= \tau_{-1}$ is read and used to determine whether the product is
formed. In other words, this can be viewed as a two-step channel with
the first step being the reading of the timer $a$ and the
second step being the determination of the product formation, which
maps the timer value to the product formation.

The information yield of the first step is given by the mutual
information between the input $\xi$ and the timer (or age) $a$, in which
we assume $\xi$ is uniformly distributed. In the case of $k_{-1}=1$
and $q_{-1}=2$, the mutual information is given by $\mathcal{I}(\xi;
a) \approx 0.078$.

\begin{figure}[htb]
    \centering
    \includegraphics[width=2.8in]{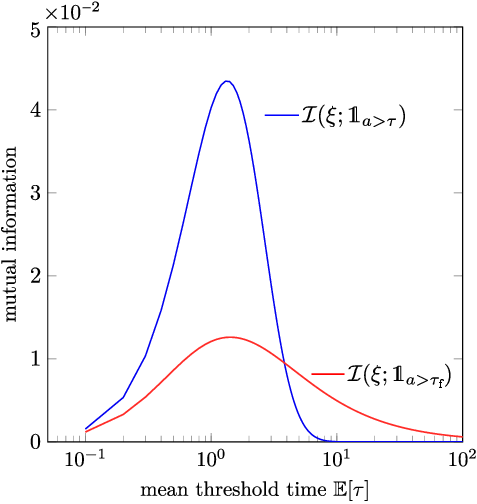}
    \vspace{4mm}
    \caption{\label{fig:mutual-information-binary} The mutual
      information between the input $\xi$ and the final output for the
      KPR and MM schemes
      ($\mathcal{I}({\xi;\mathds{1}_{a>\tau})},~\mathcal{I}(\xi;
      \mathds{1}_{a>\tau_{\rm f}}) $) in a single unbinding event, for
      different mean proofreading time. We set $k_{-1}=1$ and
      $q_{-1}=2$.}
\end{figure}

The second step involves a mapping from the timer value to the product
formation. In the case of Scheme~\eqref{eq:scheme1}, the mapping is
deterministic with a threshold $\tau$, given by
$\mathds{1}_{a>\tau}$. In the case of Scheme~\eqref{eq:scheme2}, the
mapping is stochastic with an additional exponential waiting time
$\tau_{\rm f}$ with rate $k_{\rm f}$, given by $\mathds{1}_{a>
  \tau_{\rm f}}$. We can expect that because of the additional layer
of stochasticity, the information yield of the MM scheme is much lower
than that of the KPR scheme, as confirmed by the numerical results in
Fig.~\ref{fig:mutual-information-binary}.
%
We also note that the information yield per binding-unbinding cycle is
low. Thus, multiple rounds of binding and unbinding are required to
achieve a high channel capacity.

\section{Multistep binding model}
\label{appendix:multistep_KPR}
In the main text, we have considered the analysis of the multistep
limit of the classical KPR model, where the processing time from
initial binding to full activation is taken to be a deterministic time
$\tau$.  In the classical KPR model, the processing time is subject to
noise due to finite number of steps in the activation process.

In this section, we fix the number of activation steps $m=6$ and
perform simulations using the same parameters as in the main text. The
biophysical rationale derives from the six phosphorylation sites in
the associated CD3 $\zeta$ chains. In particular, the phosphorylation
kinetics of these sites are independent of the antigen types, except
for two extra phosphorylation sites on ZAP70 that are regulated
depending on the antigen types, which we do not explicitly consider in
this paper.
\begin{figure}[htb]
    \centering
    \includegraphics[width=4.5in]{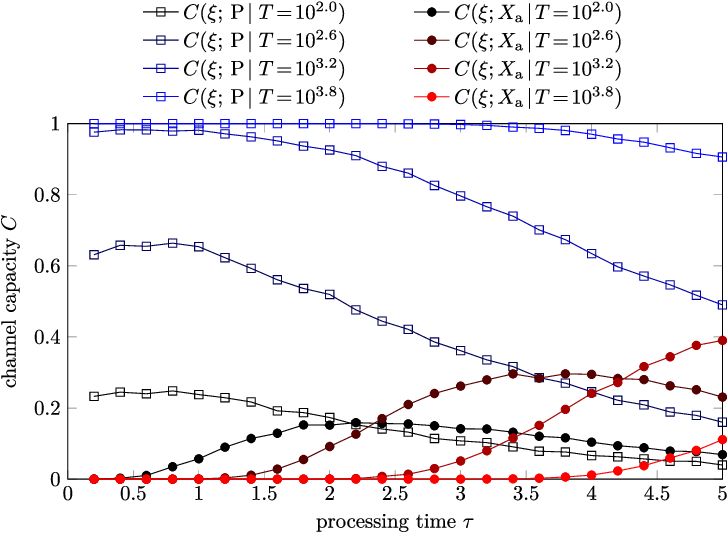}
    \vspace{4mm}
    \caption{\label{fig:multiround_multistep_simulation_CC_FLD_T} The
      channel capacities of product concentration (blue squares) and
      first activation times (red dots) as a function of processing
      time $\tau$ for various cell contact time $T$ in a six-step
      multistep binding model.  The parameter values are those used in
      Fig.~\ref{fig:channel-capacity-first-passage-time-2}.}
\end{figure}
Figure~\ref{fig:multiround_multistep_simulation_CC_FLD_T} shows a
qualitatively similar dependence of channel capacity on processing
time $\tau$ as the deterministic limit in
Fig.~\ref{fig:channel-capacity-first-passage-time-2}. The channel
capacity of product concentration is maximized at $\tau \approx
0.5/k_{1}$, irrespective of $T$, while the optimal processing time for
the channel capacity of first activation time increases with cell
contact time $T$.
\begin{figure}[htb]
    \centering
    \includegraphics[width=4.5in]{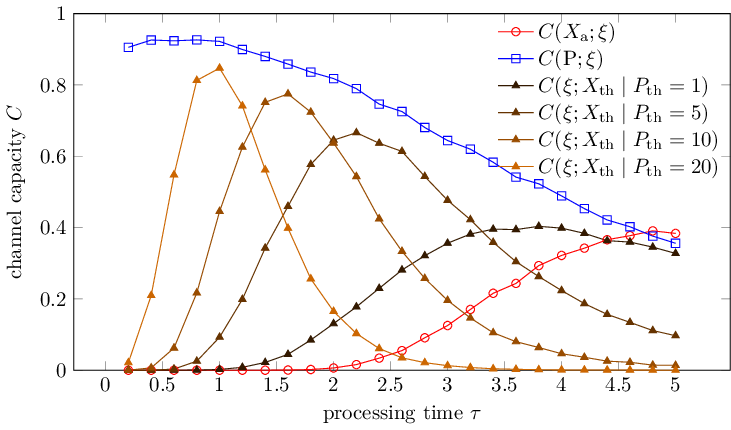}
    \vspace{4mm}
    \caption{\label{fig:multiround_multistep_simulation_CC_FLD_tau}The
      channel capacity between the input $\xi$ and the output $X_{\rm
        a}$, ${\rm P}(T)$, or $X_{\rm th}$ as a function of processing
      time $\tau$. The parameters are kept to be the same as in
      Fig.~\ref{fig:cc-estimates}. 10,000 independent Gillespie
      simulations are conducted for each $\tau$.}
\end{figure}

Similar relationships are observed for the channel capacity of
decision based on whether the protein number reaches a threshold ${\rm
  P}_{\rm th}$ at time $T$, as shown in
Fig.~\ref{fig:multiround_multistep_simulation_CC_FLD_tau} compared to
Fig.~\ref{fig:cc-estimates}.

However, in both cases, the channel capacity of
first-passage-time-based decision with 6-step activation is lower than
that of the deterministic limit, which is a consequence of the noise
in the activation process. The channel capacity of product
concentration is similar to that of the deterministic limit, which may
be explained by the buffering effect of product generation steps to
upstream noises. In particular, decrease of $C(X_{\rm th} \mid {\rm
  P}_{\rm th})$ is dependent on ${\rm P}_{\rm th}$. Larger ${\rm
  P}_{\rm th}$ leads to a smaller decrease of channel capacity. The
optimal processing time for first-passage-time-based discrimination
also increases, compared to the deterministic limit.  A quantitative
analysis awaits further investigation.

\clearpage
\setcounter{figure}{0}
\renewcommand{\thefigure}{S\arabic{figure}}
\section{Supplementary figures}
\begin{figure}[htb]
  \centering
  \includegraphics[width=4.6in]{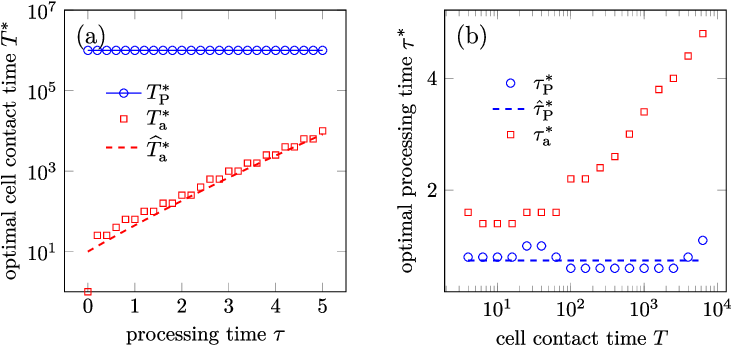}
  \vspace{4mm}
  \caption{Dependence of optimal contact times and processing times on
    each other. Symbols represent \added{results from simulations over
      a total time horizon of $10^6$, while the dashed curves represent
      analytic approximations.}  (a) Dependence of optimal cell
    contact times $T_{\rm P}^*$ and $T_{\rm a}^*$ on processing time
    $\tau$ under product-based and FPT-based strategies. \added{Since
      $C(\xi;P)$ is nondecreasing with respect to $T$, the maximizing
      $T_{\rm P}^*$ value is given as the upper limit of the
      simulation period. The approximation for $T_{\rm a}^*$,
      $\hat{T}_{\rm a}^{*} =N^{*}/K_{1}$, is given by
      Eq.~\eqref{eq_acc:N_star}.}  (b) Dependence of optimal
    processing times $\tau_{\rm P}^*$ and $\tau_{\rm a}^*$ on cell
    contact time $T$.  \added{The analytic approximation
      $\hat{\tau}_{\rm P}^{*}$ is given by Eq.~\eqref{eq:tau_star}.}
    The parameters used are the same as those in
    Fig.~\ref{fig:channel-capacity-first-passage-time-2}.}
  \label{fig:channel-capacity-first-passage-time-3}
\end{figure}
\begin{figure}[t]
  \centering
  \includegraphics[width=4.5in]{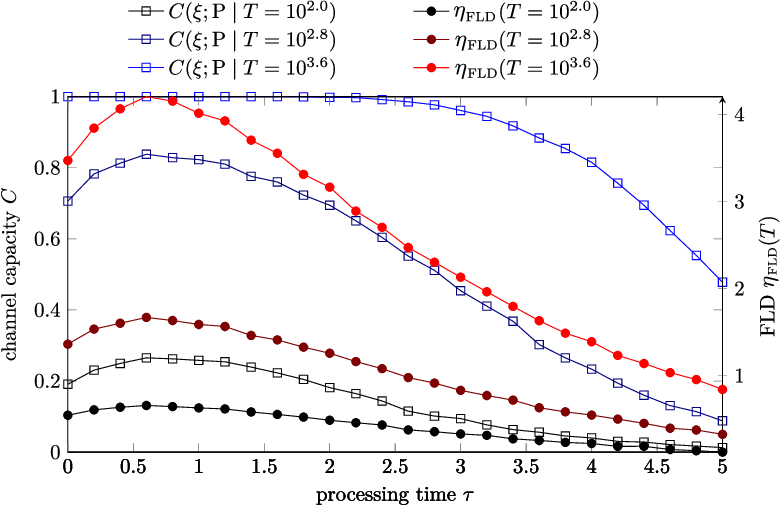}
  \vspace{4mm}
  \caption{{Dependence of the channel capacity between input $k_{-1}$ or
  $q_{-1}$ and the product ${\rm P}$ at time $T$ and the Fisher
  linear discriminant on the processing time $\tau$ for different
  values of $T$.} Both quantities are evaluated via numerical
  simulations from $10^4$ trajectories. We assumed $k_{1}=q_1 =
  0.1$, $k_{-1}=k_{-1}^* = 1$, $q_{-1}=q_{-1}^* = 2$, $\tau = 3$,
  and $k_{\rm p} = 1$ for a fast product formation rate.}
  \label{fig:multiround_simulation_CC_FLD_tau}
\end{figure}
\begin{figure}[t]
  \centering
  \includegraphics[width=4.5in]{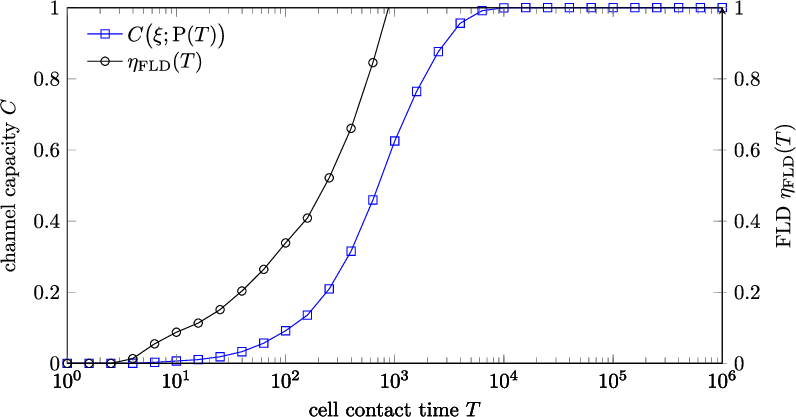}
  \vspace{4mm}
  \caption{\label{fig:multiround_simulation_CC_FLD_T} Dependence of
  the channel capacity between input $\xi$ and the product ${\rm P}$
  and the Fisher linear discriminant on the cell-cell contact time
  $T$. Both quantities are evaluated via numerical simulations from
  $10^4$ trajectories. We assumed $k_{1}=q_1 = 0.1$, $k_{-1}=k_{-1}^*
  = 1$, $q_{-1}=q_{-1}^* = 2$, $\tau = 3$, and $k_{\rm p} = 1$ for a
  fast product formation rate. For $\eta_{\rm FLD}$, when $T> 10^3$,
  their values are too high to be shown in the figure. We cropped the
  values of $\eta_{\rm FLD}$ to $(0,1)$ for better visualization.}
\end{figure}
\begin{figure}[b]
  \centering
  \includegraphics[width=5.6in]{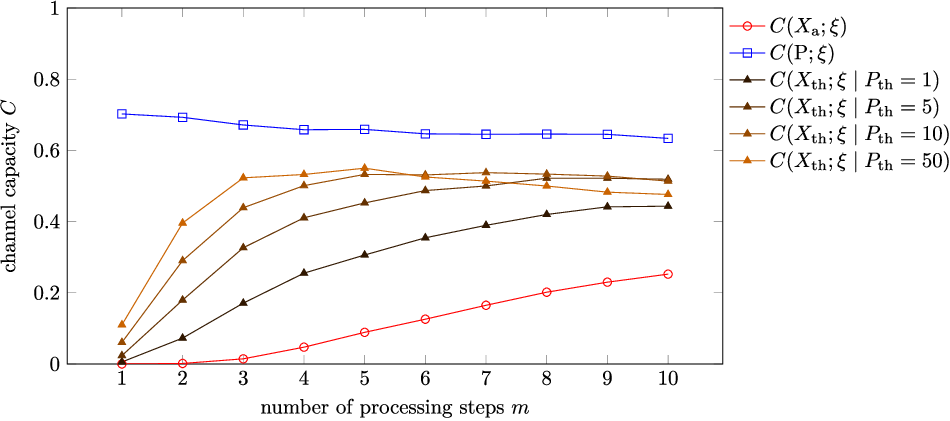}
  \vspace{4mm}
  \caption{\label{fig:multiround_multistep_simulation_CC_FLD_T_1000}Dependence
    of the channel capacity between input $\xi$ and different outputs
    on the number of proofreading steps $m$. They are obtained from
    numerical simulations of the multistep binding model with $10^5$
    trajectories for each set of parameters. We assumed $k_{1}=q_1 =
    0.1$, $k_{-1}=k_{-1}^* = 1$, $q_{-1}=q_{-1}^* = 2$, $\tau = 3$,
    and $k_{\rm p} = 1$. $T$ is fixed at $1000$. The cross-over between ${\rm P}_{\rm th}=50$ 
    and ${\rm P}_{\rm th}=10$ is a result of finite sample stochasticity.}
\end{figure}

\clearpage 

%
%
%





\end{document}